\documentclass[twocolumn,resetfootnote]{aastex701}
\shorttitle{Galactic-Scale Distribution of $^{10}$Be}
\shortauthors{Fujimoto et al.}
\accepted{2026 October 24}

\usepackage{amsmath}
\usepackage{CJK}

\begin{document}
\begin{CJK*}{UTF8}{ipxm}

\title{Galactic-Scale Distribution of $^{10}$Be and Its Implications for the Early Solar System $^{10}$Be/$^{9}$Be}

\correspondingauthor{Yusuke Fujimoto}

\author[orcid=0000-0002-2107-1460,gname=Yusuke,sname=Fujimoto]{Yusuke Fujimoto (藤本裕輔)}
\affiliation{Department of Computer Science and Engineering, University of Aizu, Aizu-Wakamatsu, Fukushima 965-8580, Japan}
\email[show]{fujimoto@u-aizu.ac.jp}

\author[orcid=0000-0002-1571-0836,gname=Steven,sname=Desch]{Steven J. Desch}
\affiliation{School of Earth and Space Exploration, Arizona State University, PO Box 876004, Tempe, AZ 85287-6004, USA}
\email{steve.desch@asu.edu}

\author[orcid=0000-0003-3383-2279,gname=Jiro,sname=Shimoda]{Jiro Shimoda}
\affiliation{Institute for Cosmic Ray Research, The University of Tokyo, 5-1-5 Kashiwanoha, Kashiwa, Chiba 277-8582, Japan}
\email{jshimoda@icrr.u-tokyo.ac.jp}

\author[orcid=0000-0003-4366-6518,gname=Shu-ichiro,sname=Inutsuka]{Shu-ichiro Inutsuka}
\affiliation{Department of Physics, Graduate School of Science, Nagoya University, Furo-cho, Chikusa-ku, Nagoya, Aichi 464-8692, Japan}
\email{inutsuka@nagoya-u.jp}

\begin{abstract}

Meteoritic inclusions indicate that the early Solar System contained a substantial abundance of the short-lived radionuclide ${}^{10}$Be. Whether it was inherited from the parent molecular cloud or requires local irradiation or late injection remains debated. Here we present the first three-dimensional simulations of a Milky Way--like disc that follow the multi-phase interstellar medium, cosmic rays (CRs), and the production and transport of ${}^{10}$Be by galactic CRs (GCRs) and in supernova remnants (SNRs). Nearly all ${}^{10}$Be is produced by GCR interactions; SNRs contribute at most about ten percent of the ${}^{10}$Be inherited by newly formed stars. The ${}^{10}$Be/${}^{9}$Be ratio traces the CR energy density, declining smoothly with galactocentric radius, and because stars form where supernovae inject CRs, their natal gas is CR-enriched roughly fivefold relative to the mean interstellar medium, which one-zone estimates miss. Accounting for a possibly higher Galactic star formation rate 4.6~Gyr ago, inheritance of GCR-produced ${}^{10}$Be can then explain the early Solar System's ${}^{10}$Be/${}^{9}$Be in our models, without late injection or local irradiation. Even when GCR-induced ${}^{10}$Be production in dense gas is reduced tenfold, the meteoritic value is reached in the innermost disc and missed by only a factor of a few elsewhere, so the main conclusion is robust to uncertain low-energy GCR penetration. The inferred birth radius of the Sun depends strongly on the CR diffusivity, ranging between $3$ and $8$~kpc, suggesting that improved modelling of low-energy CR transport could sharpen constraints on the birthplace of the Solar System.

\end{abstract}



\section{Introduction}
\label{sec:introduction}

Analyses of primitive inclusions in meteorites, such as calcium-aluminium-rich inclusions (CAIs) and chondrules, have revealed that the early Solar System was enriched in more than ten short-lived radionuclides \citep[SLRs; see the reviews by][]{LugaroEtAl2018,DeschEtAl2023}. The short half-lives ($\lesssim 10$ Myr) and diverse production sites of these nuclides provide key information about the environment in which the Solar System formed. For example, $^{26}$Al, with a half-life of 0.72 Myr, is produced by stellar nucleosynthesis in massive stars and released into the interstellar medium (ISM) via stellar winds and supernova explosions. The initial $^{26}$Al/$^{27}$Al ratio in the early Solar System thus constrains the relationship between the Solar birth environment and recent massive-star activity, including the enrichment of the parent molecular cloud by massive-star ejecta \citep[e.g.,][]{LeeEtAl1976,CameronTruran1977,JacobsenEtAl2008,BossEtAl2008,DeschEtAl2023}. The large amount of $^{26}$Al incorporated into the proto--Solar System is also thought to have served as a major radioactive heat source responsible for the melting and differentiation of asteroids and planetesimals \citep[e.g.,][]{Urey1955, GrimmMcSween1993}. Furthermore, in the present-day Milky Way, gamma-ray emission from $^{26}$Al associated with massive star-forming regions demonstrates a universal link between star formation and SLR production \citep[e.g.,][]{DiehlEtAl2006, BouchetEtAl2015, PleintingerEtAl2019}.

Another key SLR, $^{60}$Fe, has a half-life of 2.61 Myr \citep{RugelFaestermannKorschinek2026}. Because $^{60}$Fe is almost exclusively produced in supernova explosions, it provides even more specific clues regarding the relationship between the Solar System's birth environment and nearby supernova activity \citep[e.g.,][]{GounelleMeynet2012, Young2014, FujimotoEtAl2018}. In addition, traces of $^{60}$Fe have been found in deep-sea sediments and Antarctic snow, suggesting that one or more supernovae occurred near the Solar System within the last several million years \citep[e.g.,][see also \citealt{FujimotoEtAl2020b}]{KnieEtAl1999, KollEtAl2019}. These deposits also provide clues to the origin of the cosmic rays (CRs) observed near the Earth today \citep[see, e.g.,][for a recent study]{ShimodaAsanoInutsuka2026}. These examples illustrate that SLRs serve as powerful tracers of past, recent, and ongoing astrophysical events in the Galaxy.

Among the SLRs, $^{10}$Be is unique. It has a half-life of 1.387 Myr and, unlike $^{26}$Al or $^{60}$Fe, it is not a product of stellar nucleosynthesis but instead is synthesized non-thermally through spallation reactions induced by energetic cosmic rays. An early hypothesis proposed that $^{10}$Be in the proto--Solar System was produced in situ by irradiation of the solar nebula with energetic particles from the proto-Sun \citep[``irradiation scenario''; e.g.,][]{GounelleEtAl2001, Jacquet2019}. Since spallogenic production is highly sensitive to the particle flux, which decreases with distance from the proto-Sun, irradiation models generally predict spatially and temporally heterogeneous distributions of $^{10}$Be within the solar nebula. However, measurements of the initial $^{10}$Be/$^{9}$Be ratios in almost 60 CAIs (excluding the rare FUN CAIs, which show Fractionation and Unknown Nuclear isotopic effects) have shown remarkable uniformity, with values clustering around $^{10}$Be/$^{9}$Be $\approx 7\times10^{-4}$ and a scatter consistent with measurement uncertainties \citep{DunhamEtAl2022, DeschEtAl2026}\footnote{This meteoritic ratio refers to Be incorporated into solids as ordinary matter, and should not be confused with the $^{10}$Be/$^{9}$Be flux ratio of energetic cosmic-ray nuclei, which is much larger and is widely used as a radioactive clock to constrain the Galactic CR residence time and diffusive-halo size \citep[e.g.,][]{EvoliEtAl2020}.}, which strongly disfavours localized irradiation.

Alternative scenarios have been proposed in which $^{10}$Be, or the cosmic rays that produce it, were externally supplied to the presolar material (``injection scenario''): either by the magnetic trapping of galactic cosmic rays (GCRs), including the $^{10}$Be nuclei they carry, in the collapsing molecular cloud core \citep{DeschEtAl2004}, or by the irradiation of the parent molecular cloud by locally enhanced cosmic rays from a nearby supernova remnant (SNR) or from supernovae (SNe) in a superbubble \citep{TatischeffEtAl2014}. In the ``inheritance scenario'', by contrast, $^{10}$Be was already present in the parent molecular cloud before the Sun's birth and was simply inherited \citep{DunhamEtAl2022, DeschEtAl2023, DeschEtAl2026}. All of these models require consideration of multiple $^{10}$Be production processes: (1) production within SNRs via trapped cosmic rays, (2) production by cosmic rays escaping from SNRs and irradiating surrounding interstellar gas, and (3) continuous production by GCRs throughout the Galactic disc. Although the nearly uniform $^{10}$Be/$^{9}$Be ratio recorded by meteoritic inclusions favours inheritance from the parent molecular cloud, this interpretation has historically faced a quantitative difficulty: calculations based on the GCR flux in the Solar neighbourhood predicted molecular-cloud ratios of only $\sim 10^{-4}$ \citep{GounelleEtAl2001,TatischeffEtAl2014}, well below the meteoritic value. In particular, \citet{TatischeffEtAl2014} obtained a steady-state ratio of $1.3\times10^{-4}$ from background GCR irradiation and therefore invoked additional CRs from multiple SNe in a superbubble or from a nearby SNR. More recently, \citet{DeschEtAl2023} estimated that if the Sun formed within a massive spiral arm with enhanced star formation activity, GCR production alone might reproduce the high $^{10}$Be/$^{9}$Be ratio observed in meteorites. However, such estimates remain analytic and have not yet been tested by realistic numerical simulations that model $^{10}$Be transport and distribution across the galactic disc.

In this study, we perform hydrodynamic simulations of a Milky Way--like disc that self-consistently evolve interstellar gas and cosmic rays. We calculate both the production of $^{10}$Be by GCRs and SNRs and its subsequent transport through the ISM to obtain, for the first time, the galactic-scale distribution of $^{10}$Be.

\section{Methods}
\label{sec:methods}

Our simulations are designed to reproduce the present-day Milky Way Galaxy as a reference model. One of the goals of this study is to discuss the origin of the early Solar System ${}^{10}$Be/${}^{9}$Be ratio and the environment in which the Sun formed about 4.6 Gyr ago. Ideally, this would require modelling the Milky Way at that epoch. However, we assume that the global statistical properties of the Galaxy, such as its mass, size, and overall structure, have not changed significantly over the past several Gyr. Under this assumption, a present-day Milky Way model provides a reasonable approximation for studying the large-scale production and transport of $^{10}$Be in the galactic disc. The star formation rate 4.6~Gyr ago may have differed from the present-day value; we account for this in the analysis and discussion in later sections rather than in the simulation setup.

\subsection{Initial conditions} 
\label{subsec:initial_conditions}

\begin{table*}[ht!]
	\centering
	\caption{Initial conditions of the simulated galaxy}
	\label{tab:initial_conditions}
    \footnotesize
	\begin{tabular}{lcccc}
		\hline\hline
		& Dark matter halo & Stellar disc & Stellar bulge & Gas disc \\
		\hline
		Density profile & \citet{Hernquist1990} & exponential & \citet{Hernquist1990} & exponential \\
		Galaxy structure & $M_{\mathrm{halo}} = 1.254 \times 10^{12}\ \mathrm{M_{\sun}}$ & $M_{\mathrm{disc}} = 3.437 \times 10^{10}\ \mathrm{M_{\sun}}$ & $M_{\mathrm{bulge}} = 4.297 \times 10^{9}\ \mathrm{M_{\sun}}$ & $M_{\mathrm{gas}} = 6.445 \times 10^{9}\ \mathrm{M_{\sun}}$ \\
		& $r_{\mathrm{halo}} = 16.55$ kpc & $R_{\mathrm{disc}} = 3.432$ kpc & $r_{\mathrm{bulge}} = 0.343$ kpc & $R_{\mathrm{gas}} = 3.432$ kpc \\
		& & $h_{\mathrm{disc}} = 0.1\ R_{\mathrm{disc}}$ & & $h_{\mathrm{gas}} = 0.1\ R_{\mathrm{gas}}$ \\
		& & & & $T_{\mathrm{gas}} = 10^4$ K \\
		Resolution & $m_{\mathrm{halo}} = 1.254 \times 10^5\ \mathrm{M_{\sun}}$ & $m_{\mathrm{disc}} = 3.437 \times 10^3\ \mathrm{M_{\sun}}$ & $m_{\mathrm{bulge}} = 3.437 \times 10^3\ \mathrm{M_{\sun}}$ & $\Delta x = 20$ pc ($t \leq 500$ Myr) \\
		& $N_{\mathrm{halo}} = 10^7$ & $N_{\mathrm{disc}} = 10^7$ & $N_{\mathrm{bulge}} = 1.25 \times 10^6$ & $\Delta x = 10$ pc ($t > 500$ Myr) \\
		\hline
	\end{tabular}
\end{table*}

We make use of the initial conditions for an isolated Milky Way--like galactic disc developed in the AGORA High-resolution Galaxy Simulations Comparison Project \citep{AGORA_KimEtAl2014}. The parameters used in our simulation are summarized in Table~\ref{tab:initial_conditions}. Briefly, the AGORA model provides the initial structures of the four fundamental galactic components---dark matter halo, stellar disc, stellar bulge, and gas disc---based on analytic equilibrium profiles (see also \citealt{FujimotoEtAl2020a}, who used the same initial conditions). In addition to these standard components, we also include the initial distributions of $^{10}$Be and CRs specific to this study. We describe the galaxy model and these additional components in the following, and refer readers to the papers mentioned above for further details.

The dark matter and stars are represented using collisionless particles and are initialized by stochastically drawing from analytic distribution functions. Positions of the dark matter halo particles are initialized to follow a \citet{Hernquist1990} profile as,
\begin{equation}
    \rho_{\mathrm{halo}}(r) = \frac{M_{\mathrm{halo}}}{2\pi} \frac{r_{\mathrm{halo}}}{r(r+r_{\mathrm{halo}})^3},
\end{equation}
where $M_{\mathrm{halo}}$ is the mass of the dark matter halo, $r$ is the spherical radial coordinate, and $r_{\mathrm{halo}}$ is the halo scale length.
Positions of the stellar disc particles are initialized following
\begin{equation}
    \rho_{\mathrm{disc}}(R, z) = \frac{M_{\mathrm{disc}}}{4\pi R_{\mathrm{disc}}^2 h_{\mathrm{disc}}} \exp \left( -\frac{R}{R_{\mathrm{disc}}} \right) \mathrm{sech}^2 \left( \frac{z}{h_{\mathrm{disc}}} \right),
\end{equation}
where $M_{\mathrm{disc}}$ is the mass of the stellar disc, $R_{\mathrm{disc}}$ is the scale radius, $h_{\mathrm{disc}}$ is the scale height, $R = \sqrt{x^2+y^2}$ is the cylindrical radial coordinate, and $z$ is the vertical coordinate. 
Positions of the stellar bulge particles are initialized to follow a \citet{Hernquist1990} profile as,
\begin{equation}
    \rho_{\mathrm{bulge}}(r) = \frac{M_{\mathrm{bulge}}}{2\pi} \frac{r_{\mathrm{bulge}}}{r(r+r_{\mathrm{bulge}})^3},
\end{equation}
where $M_{\mathrm{bulge}}$ is the mass of the stellar bulge, $r$ is the spherical radial coordinate, and $r_{\mathrm{bulge}}$ is the bulge scale length. 

All particles in each population have uniform masses; the particle numbers and masses are listed in Table~\ref{tab:initial_conditions}.

The initial gas distributions on the grid structure are initialized following an analytic density profile as
\begin{equation}
    \rho_{\mathrm{gas}}(R, z) = \frac{M_{\mathrm{gas}}}{4\pi R_{\mathrm{gas}}^2 h_{\mathrm{gas}}} \exp \left( -\frac{R}{R_{\mathrm{gas}}} \right) \exp \left( -\frac{|z|}{h_{\mathrm{gas}}} \right),
\end{equation}
where $M_{\mathrm{gas}}$ is the mass of the gas disc, $R_{\mathrm{gas}}$ is the scale radius, and $h_{\mathrm{gas}}$ is the scale height. The initial gas fraction is set to $f_{\mathrm{gas}} = M_{\mathrm{gas}} / (M_{\mathrm{disc}} + M_{\mathrm{gas}}) = 0.16$. The initial gas temperature in the disc is set to $T_{\mathrm{gas}} = 10^4$ K. The mass distribution of all four components (halo, disc, bulge, and gas) sets an initial rotation curve of the gas disc with circular velocity $V_{\mathrm{c}}(R = 8\ \mathrm{kpc}) \simeq 240\ \mathrm{km\ s^{-1}}$, consistent with observations \citep[e.g.,][]{ReidEtAl2019}.

In addition to the gas disc, we include a uniform, low-density gas halo that fills the entire simulation box, with initial density of $7.7 \times 10^{-7}\ \mathrm{cm}^{-3}$, zero initial velocity, and initial temperature of $10^6$ K.

The initial mass distribution of $^{10}$Be is set to follow that of the gas throughout the simulation box, assuming a uniform mass fraction of $f_{^{10}\mathrm{Be}} = 10^{-18}$ relative to the gas density,
\begin{equation}
\rho_{\mathrm{^{10}Be}}(R,z) = f_{^{10}\mathrm{Be}}\,\rho_{\mathrm{gas}}(R,z).
\end{equation}
This choice has no practical effect since the initial $^{10}$Be mass decays rapidly with the half-life of $t_{1/2} = 1.387$ Myr \citep{ChmeleffEtAl2010, KorschinekEtAl2010}, and the subsequent evolution is dominated by continuous $^{10}$Be production by cosmic rays.

The initial CR energy density is mapped proportionally to the gas distribution throughout the simulation box, such that
\begin{equation}
\epsilon_{\mathrm{CR}}(R,z) = \epsilon_{\mathrm{CR}, 0}\left[\frac{\rho_{\mathrm{gas}}(R,z)}{\rho_{\rm ref}}\right],
\end{equation}
where $\epsilon_{\mathrm{CR}, 0} = 1.0\times10^{-13}\ {\rm erg\ cm^{-3}}$ ($\simeq 0.06\ \mathrm{eV\ cm^{-3}}$) and $\rho_{\rm ref}$ corresponds to the gas mass density for a number density of $n_{\rm H}=1\ {\rm cm^{-3}}$. The generation and diffusion of CRs quickly dominate the overall CR distribution, and the total CR energy in the disc subsequently approaches a quasi-steady state that is largely insensitive to the assumed initial condition, as we will show in Section~\ref{subsec:evolution_of_the_galactic_disc}.

\subsection{Numerical methods} 
\label{subsec:numerical_methods}

Our simulations follow the evolution of a Milky Way--like galaxy using the Eulerian adaptive mesh refinement (AMR) code \textsc{Enzo} \citep{BryanEtAl2014, Brummel-SmithEtAl2019}. We employ live $N$-body representations of the dark matter halo and stellar disc using the particle-mesh dynamics implemented in \textsc{Enzo}, which allows the galactic disc to form spiral arms self-consistently through gravitational interactions among dark matter, stars, and gas.  

To follow the motion of the gas, we use the \textsc{Zeus} hydrodynamics solver as implemented in the \textsc{Enzo} code, which employs a finite-difference scheme for solving hyperbolic partial differential equations with second-order accuracy in space and first-order accuracy in time \citep{StoneNorman1992a}. To capture the thermal evolution of the gas, we employ the \textsc{Grackle} cooling library \citep{SmithEtAl2017}, which provides radiative cooling under the optically thin approximation. Cooling rates for primordial gas and metals are tabulated assuming ionization equilibrium, with the tables pre-computed using \textsc{Cloudy} \citep{FerlandEtAl2013}. An extragalactic UV background, based on the model of \citet{HaardtMadau2012}, serves as the source of photo-heating and photo-ionization in the calculation of these rates. To represent heating processes relevant to a Milky Way--like ISM, we include photoelectric heating caused by far-ultraviolet (FUV) photons that eject electrons from dust grains. A uniform heating rate of $8.5 \times 10^{-26}\ \mathrm{erg\ s^{-1}}$ per hydrogen atom is adopted throughout the simulation box. Self-gravity of the gas is also taken into account. 

We treat $^{10}\mathrm{Be}$ as a passive scalar that is advected with the gas, and that decays with the half-life of $t_{1/2} = 1.387$ Myr \citep{ChmeleffEtAl2010, KorschinekEtAl2010}, following the exponential decay law
\begin{equation}
    \rho_{\mathrm{^{10}Be}} (t + \Delta t) = \rho_{\mathrm{^{10}Be}} (t) \exp\left(- \ln{(2)} \frac{\Delta t}{t_{1/2}}\right),
\end{equation}
where $\rho_{\mathrm{^{10}Be}}(t)$ is the $^{10}\mathrm{Be}$ density of a cell at time $t$ and $\Delta t$ is the local time-step. We do not include dust grain physics because the dust grains and gas are expected to be well coupled at the spatial resolution achieved in this simulation.

The evolution of CRs is modelled using the two-fluid approximation \citep{DruryFalle1986, JunClarkeNorman1994}, as implemented in \textsc{Enzo} by \citet{SalemBryan2014}. In this model, the CRs are treated as an ultrarelativistic ideal gas representing a population of relativistic particles, with an adiabatic index of $\gamma_{\mathrm{CR}}=4/3$, dynamically coupled to the thermal plasma except for spatial diffusion. The CR component is treated as a massless fluid that carries energy and pressure but no momentum and interacts with the gas solely through CR pressure gradients in the momentum equation. This two-fluid approach assumes an isotropic (scalar) CR pressure rather than an anisotropic tensor form. The CR diffusion coefficient, $\kappa_{\mathrm{CR}}$, is assumed to be spatially isotropic and independent of the local gas properties, with a constant value of $\kappa_{\mathrm{CR}}=3\times10^{28}\ {\rm cm^{2}\ s^{-1}}$ during the relaxation phase; the values adopted in the high-resolution phase are given in Section~\ref{subsec:parameter_study}. We neglect diffusion of CRs in energy as well as energy losses due to Coulomb collisions, ionization of the ambient gas, and hadronic (pion-production) interactions; the consequences of neglecting these losses are discussed in Section~\ref{subsec:limitations}. We also neglect the motion of scattering centres relative to the bulk fluid, such as Alfv\'enic drift, whose effects are implicitly captured by the isotropic CR diffusion term.
The two-fluid model thus evolves only the total CR energy density---the CR energy budget, which is dominated by particles with kinetic energies of $\sim 1$--$10$~GeV---and retains no information on the CR spectrum. How this energy budget is coupled to the production of $^{10}$Be is specified in Section~\ref{subsec:10Be_production_model}.

\subsection{Grid structure and refinement criteria}
\label{subsec:grid_structure_and_refinement_criteria}

Our simulation adopts an adaptive mesh refinement (AMR) grid configuration and refinement strategy similar to those described by \citet{GoldbaumEtAl2015}. 
The computational domain covers a cubic box with a side length of $1.31072\ \mathrm{Mpc}$, using isolated boundary conditions for gravity and periodic boundaries for the hydrodynamic variables. 
The base grid consists of $64^3$ cells, corresponding to a base-grid cell size of $20.48\ \mathrm{kpc}$, over which we apply five additional levels of static refinement. 
Each successive level encloses a region smaller by a factor of $2$ in linear size, resulting in the galactic disc being contained within a $(40.96\ \mathrm{kpc})^3$ volume, corresponding to a cell size of 640 pc.

Beyond the static refinement levels, five further levels of adaptive refinement are employed, 
resulting in a finest spatial resolution of $\Delta x = 20$ pc during the first 500 Myr of evolution, when the maximum refinement level is set to $l = 10$. 
After 500 Myr, we increase the resolution by one additional refinement level ($l = 11$), achieving a finest cell size of $\Delta x = 10$ pc.
This two-stage refinement acts as a relaxation strategy that allows the simulated galaxy to settle into quasi-statistical equilibrium before higher-resolution structures are introduced. 
The first 500 Myr corresponds to approximately two rotation periods at a galactocentric radius of 10 kpc.

Cells are adaptively refined based on several physical criteria. 
To ensure proper resolution of the dark matter halo, a cell is refined whenever the total particle mass within it exceeds $1.72\times10^6\ \mathrm{M_{\sun}}$, equivalent to roughly 14 dark matter simulation particles. 
For the gas component, refinement is triggered when the total gas mass in a cell surpasses $2.15\times10^4\ \mathrm{M_{\sun}}$. 
To prevent artificial fragmentation, we also refine a cell when the Jeans length, $L_{\mathrm{J}} = c_{\mathrm{s}}\sqrt{\pi/(G\rho)}$, becomes smaller than 16 cell widths, which comfortably satisfies the \citet{TrueloveEtAl1997} resolution criterion. 
Additionally, zones containing star particles are always refined to the maximum level in order to capture stellar feedback processes accurately. 
Finally, to maintain the Jeans length resolved after collapse at the highest refinement level, a pressure floor is applied to guarantee that the Jeans length spans at least four cells at the maximum resolution.

\subsection{Star formation and feedback}
\label{subsec:star_formation_and_feedback}

We adopt a star formation and feedback model following the prescription described in \citet{FujimotoEtAl2018, FujimotoEtAl2020a}. 
Gas becomes eligible for star formation when its number density exceeds a threshold of $n_{\mathrm{sf}} = 13\ \mathrm{cm^{-3}}$ for $\Delta x = 20$ pc or $30\ \mathrm{cm^{-3}}$ for $\Delta x = 10$ pc.
The star formation efficiency per free-fall time for gas above the density threshold is set to 0.01 \citep[see the review by][]{KrumholzMcKeeBlandHawthorn2019}. 
A minimum star particle mass of $300\ \mathrm{M_{\sun}}$ is imposed, and star particles are formed stochastically rather than deterministically, so that not every eligible cell forms a star particle at each time-step. 
In practice, all star particles are created through this stochastic sampling, and the corresponding gas mass is removed from the host cell.

To model the ionizing luminosities and stellar winds from massive stars, as well as subsequent SN explosions, we use the \textsc{Slug} stellar population synthesis code \citep{daSilvaFumagalliKrumholz2012, daSilvaFumagalliKrumholz2014, KrumholzEtAl2015}; for each star particle, the \textsc{Slug} code stochastically draws individual stars from the initial mass function (IMF), tracks their mass- and age-dependent ionizing luminosities and stellar wind mechanical luminosities, and determines when individual stars explode as SNe. 
Using the quantities calculated by the \textsc{Slug} code, we include stellar feedback from photoionization, stellar winds, and SNe, following the prescription of \citet{GoldbaumEtAl2016}. 

Briefly, the H~\textsc{ii} region photoionization feedback is implemented by heating cells within the Str\"{o}mgren radius to $10^4$~K, where the radius is estimated at every time-step from the total ionizing luminosity of each star particle.
We include the effect of stellar winds by adding thermal energy, computed by the \textsc{Slug} code, to each star particle's host cell.
For each SN that occurs in any given time-step, we inject a total momentum of $p_{\mathrm{SN}} = 5 \times 10^5\ \mathrm{M_{\sun}\ km\ s^{-1}}$ \citep{KimOstriker2015, GentryEtAl2017}, directed radially outward into the 26 neighbouring cells.
The kinetic energy increase associated with the injected momentum is subtracted from the available energy budget of $(1-f_{\mathrm{CR}})\times10^{51}$ erg, and the remaining energy is deposited in the SN host cell and its neighbouring cells as thermal energy.
Here, $f_{\mathrm{CR}}$ denotes the fraction of an SN's total energy transferred to the relativistic CR fluid, and we set $f_{\mathrm{CR}} = 0.1$.
In contrast to the kinetic and thermal SN feedback to the gas, the CR feedback energy of $f_{\mathrm{CR}}\times10^{51}$ erg is injected into the SN host cell as a CR energy field.
We also include gas mass injection from stellar winds and SNe into each star particle's host cell at every time-step.

\subsection{\texorpdfstring{$^{10}$Be}{10Be} production model}
\label{subsec:10Be_production_model}

The short-lived radionuclide $^{10}$Be is mainly produced through spallation reactions between energetic CRs and interstellar nuclei such as C, N, and O. 
In our model, we consider two distinct mechanisms that contribute to the interstellar $^{10}$Be abundance. 
First, GCRs continuously interact with diffuse interstellar gas, providing a quasi-steady background level of $^{10}$Be production throughout the galaxy. 
Second, $^{10}$Be can also be synthesized locally within SNRs by freshly accelerated CRs before they are released into the ISM. 
These two mechanisms represent the continuous and the episodic components of $^{10}$Be production. Throughout this paper we refer to the $^{10}$Be produced by the ambient, galaxy-wide CR population---the CR field evolved in our two-fluid model---as \textit{GCR-produced}, and to that produced inside SNRs by freshly accelerated CRs, before they are released into the ISM, as \textit{SNR-produced}. 
Of the three production processes listed in Section~\ref{sec:introduction}, the first corresponds to the SNR-produced channel, while the second and third are both included in the GCR-produced channel: CRs released from SNRs become part of the CR fluid, and the resulting local enhancement of the CR energy density around recent supernovae is captured to the extent permitted by the two-fluid model and the $10$~pc resolution.
In the following subsections, we describe the formulation of each process as implemented in our simulation.

\subsubsection{GCR production of \texorpdfstring{$^{10}$Be}{10Be}}

The production rate of $^{10}$Be per hydrogen atom, $P_{^{10}\mathrm{Be}}$ [atoms yr$^{-1}$ H$^{-1}$], is calculated by integrating the product of the differential GCR intensity and the spallation cross section over particle energy:
\begin{equation}
P_{^{10}\mathrm{Be}} = \sum_{i=\mathrm{C,N,O}} \sum_{j=p,\alpha} \int_0^\infty 4\pi\, \chi_i\, J_j(E)\, \sigma_{i,j\rightarrow ^{10}\mathrm{Be}}(E)\, \mathrm{d}E .
\end{equation}
Here, $\chi_i = n_i / n_{\mathrm{H}}$ is the abundance of element $i$ relative to hydrogen in the ISM, $J_j(E)$ [cm$^{-2}$ s$^{-1}$ sr$^{-1}$ (MeV/nucleon)$^{-1}$] is the differential intensity of species $j$ (protons or $\alpha$-particles), and $\sigma_{i,j\rightarrow ^{10}\mathrm{Be}}(E)$ [cm$^2$] is the energy-dependent cross section for $^{10}$Be production from target nuclei $i$ (C, N, O).

\citet{TatischeffEtAl2014} derived a production rate of
\[
P_{^{10}\mathrm{Be,Local}} = 1.16\times10^{-21}\ \mathrm{yr^{-1}\ H^{-1}},
\]
in the Local ISM based on the GCR proton and $\alpha$-particle fluxes measured by \textit{Voyager~1} and \textit{PAMELA}.
We adopt this value as a reference and scale it linearly
with the ratio of the cosmic-ray energy density in each computational cell to that of the Local ISM:
\begin{equation}
P_{^{10}\mathrm{Be}}(\boldsymbol{x}) = P_{^{10}\mathrm{Be,Local}} \frac{\epsilon_{\mathrm{CR}}(\boldsymbol{x})}{\epsilon_{\mathrm{CR,Local}}}.
\label{eq:P_10Be}
\end{equation}
This scaling assumes that the GCR spectral shape does not vary significantly across the galactic disc, so that the $^{10}$Be production rate is proportional to the total CR energy density.
Because both CR transport and CR energy losses are in reality energy dependent, this is an approximation; we assess its consequences through the parameter study in $\kappa_{\rm CR}$ (Section~\ref{subsec:parameter_study}) and discuss its limitations in Section~\ref{subsec:limitations}.\footnote{As we will show, GCR spallation produces the vast majority of the $^{10}$Be inherited by stars in our simulations (Section~\ref{subsec:contribution_SNR_10Be}), so that Equation~\ref{eq:P_10Be} is the single most consequential assumption of this work.}
We adopt $\epsilon_{\mathrm{CR,Local}} = 1.6\times10^{-12}\ \mathrm{erg\ cm^{-3}}$ ($\simeq1.0\ \mathrm{eV\ cm^{-3}}$) as a representative cosmic-ray energy density of the Local ISM, consistent with classical estimates \citep{Webber1998} and recent measurements \citep{CummingsEtAl2016}.

The temporal evolution of the $^{10}$Be mass density in each cell is expressed as
\begin{equation}
\frac{\mathrm{d}}{\mathrm{d}t} \left( \rho_{^{10}\mathrm{Be}} \right)_{\mathrm{cell}} = m_{^{10}\mathrm{Be}}\, P_{^{10}\mathrm{Be}}\, n_{\mathrm{H}} 
\label{eq:rho10Be_evolution}
\end{equation}
where $m_{^{10}\mathrm{Be}} = 1.6625\times10^{-23}$~g is the atomic mass of $^{10}$Be, and $n_{\mathrm{H}}$ [cm$^{-3}$] is the hydrogen number density of the cell.

\subsubsection{SNR production of \texorpdfstring{$^{10}$Be}{10Be}}

In addition to the production of $^{10}$Be by GCRs, $^{10}$Be can also be synthesized inside SNRs through interactions between shock-accelerated cosmic rays and the surrounding interstellar medium.
During the Sedov--Taylor (adiabatic) phase, freshly accelerated protons and $\alpha$-particles collide with ambient nuclei such as $^{12}$C, $^{14}$N, and $^{16}$O, producing $^{10}$Be through spallation reactions.
The newly produced $^{10}$Be is then released to the ISM when the remnant enters the radiative and snowplow phase, where radiative cooling becomes dominant and previously trapped CRs begin to escape into the interstellar medium.

The local production rate of $^{10}$Be per volume is
\begin{equation}
\begin{split}
q_{^{10}\mathrm{Be}}(R,t) &=
\sum_{i=\mathrm{C,N,O}} \sum_{j=p,\alpha}
\int_0^\infty n_i(R,t)\, n_j(E,R,t)\, \\
&\qquad \times \sigma_{i,j\rightarrow ^{10}\mathrm{Be}}(E)\, v_j(E)\, \mathrm{d}E ,
\end{split}
\end{equation}
where $n_i(R,t)$ [cm$^{-3}$] is the number density of target nuclei $i$ in the shocked region, $n_j(E,R,t)$ [cm$^{-3}$ (MeV/nucleon)$^{-1}$] is the differential number density of accelerated species $j$, $\sigma_{i,j\rightarrow ^{10}\mathrm{Be}}(E)$ [cm$^2$] is the spallation cross section, and $v_j(E)$ is the particle speed.

The total production rate within the remnant is obtained by volume integration,
\begin{equation}
Q_{^{10}\mathrm{Be}}(t) = \int_0^{R_{\mathrm s}(t)} q_{^{10}\mathrm{Be}}(R,t)\, 4\pi R^2\, \mathrm{d}R ,
\end{equation}
where $R_{\mathrm s}(t)$ is the shock radius. During most of the Sedov--Taylor phase, $Q_{^{10}\mathrm{Be}}$ scales with the ambient hydrogen density as
\begin{equation}
\frac{Q_{^{10}\mathrm{Be}}}{n_{\mathrm H}} \simeq (3\text{--}4)\times10^{33}\ \mathrm{cm^{3}\ s^{-1}} ,
\end{equation}
as shown by \citet{TatischeffEtAl2014}.
In this work we adopt the median value $C_{^{10}\mathrm{Be}}=3.5\times10^{33}\ \mathrm{cm^{3}\ s^{-1}}$ and write
\begin{equation}
Q_{^{10}\mathrm{Be}} \simeq C_{^{10}\mathrm{Be}}\, n_{\mathrm H} .
\end{equation}

The total $^{10}$Be mass synthesized during the Sedov--Taylor phase is
\begin{equation}
M_{^{10}\mathrm{Be}} = m_{^{10}\mathrm{Be}} \int_{t_{\mathrm{ST}}}^{t_{\mathrm{rad}}} Q_{^{10}\mathrm{Be}}(t)\, \mathrm{d}t ,
\end{equation}
where $t_{\mathrm{ST}}$ is the free-expansion to Sedov--Taylor transition time and $t_{\mathrm{rad}}$ is the onset time of the radiative, pressure-driven snowplow phase.
Following \citet[][]{BlondinEtAl1998}, the latter is estimated as
\begin{equation}
t_{\mathrm{rad}} \approx (2.9\times10^4~\mathrm{yr})
\left(\frac{E_{\mathrm{SN}}}{10^{51}\ \mathrm{erg}}\right)^{4/17}
\left(\frac{n_{\mathrm H}}{1\ \mathrm{cm^{-3}}}\right)^{-9/17} .
\end{equation}
Typically $t_{\mathrm{ST}}\sim10^3$~yr and $t_{\mathrm{rad}}\sim10^4$~yr, so $t_{\mathrm{rad}}-t_{\mathrm{ST}}\approx t_{\mathrm{rad}}$ in practice.
In addition, because $t_{\mathrm{rad}}\ll t_{1/2}$, radioactive decay of $^{10}$Be inside the SNR is neglected.

In our simulations each supernova injects $^{10}$Be locally as a source term in the same step as the momentum and energy feedback described in Section~\ref{subsec:star_formation_and_feedback}.
Specifically we add to the host cell a mass
\begin{equation}
M_{^{10}\mathrm{Be}} = m_{^{10}\mathrm{Be}}\, C_{^{10}\mathrm{Be}}\, n_{\mathrm H}\, t_{\mathrm{rad}} ,
\end{equation}
where $n_{\mathrm H}$ is evaluated as the volume-weighted mean hydrogen number density of the surrounding cells to represent the local ambient medium.

\subsubsection{Treatment of the stable isotope \texorpdfstring{$^{9}$Be}{9Be}}
\label{subsubsec:9Be_treatment}
 
Our simulations follow the production, transport, and decay of $^{10}$Be, but do not solve for the stable isotope $^{9}$Be. The two isotopes require fundamentally different treatments: $^{10}$Be, with $t_{1/2} = 1.387$~Myr, attains a quasi-steady abundance on its $\sim$Myr decay timescale, short compared with the simulation time, whereas $^{9}$Be is stable and has accumulated over the entire CR-irradiation history of the Galaxy, so its abundance cannot be obtained from a simulation spanning only a few galactic rotation periods. We therefore impose the $^{9}$Be abundance in post-processing when constructing the $^{10}$Be/$^{9}$Be ratio (Section~\ref{subsec:10Be_9Be_distribution}). As the simplest assumption, we take the $^{9}$Be number density to scale with the hydrogen number density---so that $^{9}$Be is, by construction, perfectly mixed with the gas---with a spatially uniform abundance fixed to the Solar photospheric value, $^{9}$Be/H $= 2.4\times10^{-11}$ \citep{AsplundEtAl2009}. This is essentially the treatment adopted by \citet{TatischeffEtAl2014}, whose local $^{10}$Be production rate provides our normalization (Equation~\ref{eq:P_10Be}). It is also consistent with the cosmic-time budget of Be production: $^{9}$Be is produced by the same CR spallation of ambient CNO nuclei as $^{10}$Be, and galactic chemical evolution models of this process, integrated over the $\sim 10$~Gyr history of the disc, reproduce a present-day Be/H of order $10^{-11}$ \citep[e.g.,][]{Prantzos2012}. A spatially uniform $^{9}$Be/H is nevertheless an idealization: because both the CR flux and the CNO target abundance are higher in the inner disc, a negative radial gradient of $^{9}$Be/H is physically plausible. The choice of normalization for the comparison with the meteoritic value---including the small difference between the photospheric and protosolar abundances---is discussed in Section~\ref{subsec:initial_10Be_9Be_solar_birth_environment}, the effect of a radial $^{9}$Be gradient is examined in Section~\ref{subsec:inherited_10Be_9Be_across_disc}, and the consequences of the uniform assumption for the inner disc are discussed in Section~\ref{subsec:limitations}.

\subsection{Parameter study}
\label{subsec:parameter_study}

\begin{table*}[ht]
\centering
\caption{Summary of parameter variations explored in this study. $\kappa_{\rm CR}$ is the CR diffusion coefficient (Section~\ref{subsec:numerical_methods}) and $P_{\mathrm{^{10}Be}}$ the GCR production rate of $^{10}$Be (Section~\ref{subsec:10Be_production_model}); ``Phase-independent'' means that $P_{\mathrm{^{10}Be}}$ is applied uniformly regardless of the ISM phase, whereas in Model~4 it is reduced by a factor of ten in dense gas.}
\label{tab:parameters}
\begin{tabular}{llll}
\hline
\hline
Simulation stage & Model & $\kappa_{\rm CR}$ & $P_{\mathrm{^{10}Be}}$ \\
\hline
$t \leq 500\ \mathrm{Myr}$ & All     & High ($3\times10^{28}\ \mathrm{cm^{2}\ s^{-1}}$) & Phase-independent \\
\hline
$t > 500\ \mathrm{Myr}$ & Model 1 & High   ($3\times10^{28}\ \mathrm{cm^{2}\ s^{-1}}$) & Phase-independent \\
                        & Model 2 & Middle ($1\times10^{28}\ \mathrm{cm^{2}\ s^{-1}}$) & Phase-independent \\
                        & Model 3 & Low    ($3\times10^{27}\ \mathrm{cm^{2}\ s^{-1}}$) & Phase-independent \\
                        & Model 4 & Middle ($1\times10^{28}\ \mathrm{cm^{2}\ s^{-1}}$) & $1/10$ for $n_{\rm H} > 100\ \mathrm{cm^{-3}}$ \\
\hline
\end{tabular}
\end{table*}

As described in Section~\ref{subsec:grid_structure_and_refinement_criteria}, the simulation was first evolved for 500~Myr with a spatial resolution of $\Delta x = 20$~pc.
After this phase, the resolution was increased to $\Delta x = 10$~pc for the subsequent evolution to resolve finer ISM structures and CR distributions.
During the initial 500~Myr, we ran a single model to establish a quasi-steady galactic disc.
Beyond 500~Myr, we performed a set of higher-resolution runs to investigate the dependence of the results on key physical parameters. The following two parameters were systematically varied.

\textbf{CR diffusion coefficient:}
Because $^{10}$Be production is dominated by sub-GeV protons and $\alpha$-particles, the relevant diffusion coefficient, $\kappa_{\rm CR}$, should reflect the transport properties of cosmic rays with energies of $\sim100$--$300\ \mathrm{MeV}$.
Galactic CR propagation studies infer diffusion coefficients of $\sim 10^{28}$ to a few $\times10^{28}\ \mathrm{cm^2\ s^{-1}}$ at GeV energies \citep[e.g.,][]{YanasakEtAl2001, EvoliEtAl2020}, while the energy dependence of CR transport and its variation among the phases of the multiphase ISM remain uncertain \citep[e.g.,][]{AmatoCasanova2021, ArmillottaOstrikerJiang2022}.
We therefore treat $\kappa_{\rm CR}$ as a constant free parameter of the model and consider a plausible range of $3\times10^{27}$--$3\times10^{28}\ \mathrm{cm^2\ s^{-1}}$, spanning one order of magnitude in the uncertain diffusivity of the sub-GeV CRs most relevant to $^{10}$Be production.
During the early evolution of the simulated galaxy ($t \leq 500\ \mathrm{Myr}$), we adopted a fixed diffusion coefficient of $\kappa_{\mathrm{CR}} = 3\times10^{28}\ \mathrm{cm^2\ s^{-1}}$.
This choice is motivated by the fact that CRs with energies of order $\sim1\ \mathrm{GeV}$ dominate the overall CR pressure in the ISM, so that the build-up of a quasi-steady galactic disc at early times is primarily regulated by GeV-scale CR transport.
For $t>500\ \mathrm{Myr}$, in the high-resolution phase, we explored three values within this range: Models~1, 2, and 3 adopt $\kappa_{\mathrm{CR}} = 3\times10^{28}$, $1\times10^{28}$, and $3\times10^{27}\ \mathrm{cm^2\ s^{-1}}$, respectively.

\textbf{GCR production rate of $^{10}$Be:} The $^{10}$Be production rate per hydrogen atom in each computational cell, $P_{\mathrm{^{10}Be}}(\boldsymbol{x})$, is computed by scaling the locally observed production rate \citep{TatischeffEtAl2014} with the ratio of the cosmic-ray energy density in the cell to that of the Local ISM (Equation~\ref{eq:P_10Be}). Thus, $P_{\mathrm{^{10}Be}}$ varies spatially according to the local CR energy density, but it does not include any additional dependence on the ISM phase (e.g., molecular, atomic, or ionized gas). In this sense, Models~1--3 assume that the GCR irradiation responsible for $^{10}$Be production is independent of the ISM phase. However, several studies suggest that the flux of low-energy cosmic rays, which dominate $^{10}$Be production, may be significantly attenuated inside dense molecular clouds. Early theoretical work \citep{SkillingStrong1976, CesarskyVolk1978} argued that sub-GeV cosmic rays ($E \lesssim 300\ \mathrm{MeV}$) can be efficiently excluded from clouds due to strong ionization losses. More recently, \citet{EverettZweibel2011} demonstrated that the breakdown of the cosmic-ray streaming instability and strong ion--neutral damping at cloud boundaries reduces the cosmic-ray pressure inside clouds, typically by only a modest amount (7.5\%) but potentially by up to an order of magnitude under certain physical conditions. We do not attempt to model these effects in Models~1--3: following \citet{TatischeffEtAl2014}, we retain the assumption that low-energy GCRs penetrate dense clouds freely. Because the $^{10}$Be production rate scales with the cosmic-ray energy density, stronger attenuation in dense clouds would imply a corresponding reduction in the local $^{10}$Be yield. To examine this possibility and assess its impact on the galactic-scale $^{10}$Be distribution, we performed an additional parameter run in which $P_{\mathrm{^{10}Be}}$ was reduced by a factor of ten for gas cells with hydrogen number density $n_{\rm H} > 100\ \mathrm{cm^{-3}}$. The factor-of-ten reduction is chosen to probe an extreme suppression of the local $^{10}$Be production rate, motivated qualitatively by the large decreases in CR pressure or density found in extreme regions of the parameter space explored by \citet{EverettZweibel2011}; Model~4 is therefore intended as a deliberately strong sensitivity test rather than an equally plausible alternative to Models~1--3.

These parameter variations (summarized in Table~\ref{tab:parameters}) allow us to assess the impact of CR transport and dense-gas production suppression on the global distribution and abundance of ${}^{10}$Be in the simulated galaxy.

\section{Results} 
\label{sec:results}

\subsection{Evolution of the galactic disc} 
\label{subsec:evolution_of_the_galactic_disc}

\begin{figure}
\centering
\includegraphics[width=\columnwidth]{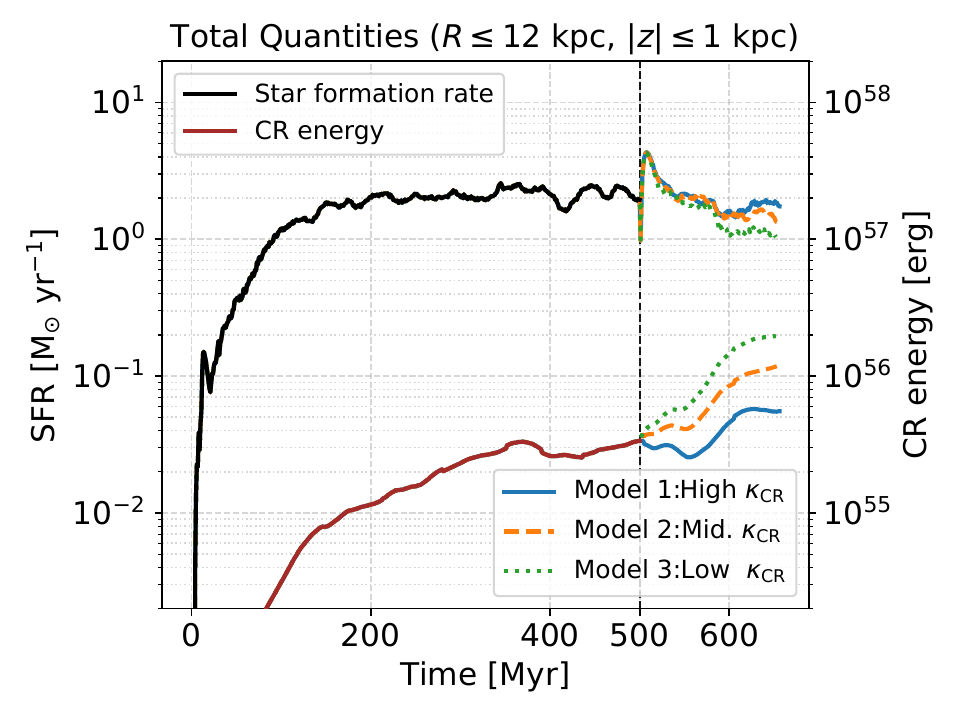}
\caption{
Time evolution of the total star formation rate (SFR; black solid line) and total CR energy (brown solid line) within the galactic disc ($R \leq 12$~kpc, $|z| \leq 1$~kpc), covering the low-resolution relaxation phase ($t \leq 500$~Myr) and the high-resolution phase ($t > 500$~Myr). The vertical black dashed line marks $t = 500$~Myr, the end of the initial relaxation phase, at which point the resolution is increased from 20~pc to 10~pc. During the relaxation phase, a single run is evolved. In the high-resolution phase, the CR energy is shown separately for Model~1 (High~$\kappa_{\rm CR}$; blue solid), Model~2 (Middle~$\kappa_{\rm CR}$; orange dashed), and Model~3 (Low~$\kappa_{\rm CR}$; green dotted). Model~4, which differs from Model~2 only in the GCR production rate of $^{10}$Be in dense gas, is not shown as it yields the same SFR and CR energy as Model~2.
}
\label{fig:time_evolution_SFR_CR}
\end{figure}

We evolve the simulation toward a quasi-steady state, treating the initial 500 Myr as a relaxation phase. Once the disc reaches a quasi-equilibrium, the maximum resolution is increased from 20 pc to 10 pc for the main four runs used in our analysis.
Figure~\ref{fig:time_evolution_SFR_CR} presents the time evolution of the total star formation rate (SFR) and the total CR energy within the region of $R \leq 12\ \mathrm{kpc}$ and $|z| \leq 1\ \mathrm{kpc}$, covering both the low-resolution relaxation phase ($t \leq 500$~Myr) and the subsequent high-resolution phase ($t > 500$~Myr), during which the four runs are evolved. For clarity, the high-resolution SFR and CR energy are shown only for Models~1, 2, and 3, which differ in the CR diffusion coefficient $\kappa_{\rm CR}$; Model~4, which modifies only the GCR production rate of $^{10}$Be in dense gas, is omitted here as it yields the same SFR and CR energy as Model~2. After $t \sim 200\ \mathrm{Myr}$---roughly one orbital period at $R = 8\ \mathrm{kpc}$---the SFR settles to a nearly constant level of $\sim 2\ M_{\odot}\ \mathrm{yr^{-1}}$, which is broadly consistent with the current SFR of the Milky Way \citep[e.g.,][]{ChomiukPovich2011, LicquiaNewman2015}.
The total CR energy, by contrast, continues to rise beyond $t \sim 200\ \mathrm{Myr}$ as supernovae steadily inject CRs, but its growth slows as CR escape increasingly balances injection, approaching a quasi-saturated value of a few $10^{55}\ \mathrm{erg}$ after $t \sim 400\ \mathrm{Myr}$.
Considering that the SFR has stabilized and the growth of the CR energy has significantly slowed by $t \sim 500\ \mathrm{Myr}$, we regard the galactic disc as having reached a quasi-equilibrium state in the initial low-resolution stage.

After the resolution is increased at $t = 500$~Myr, the SFR briefly drops and then rebounds before relaxing back to its pre-refinement level. These refinement steps are visible in Figure~\ref{fig:time_evolution_SFR_CR} as sudden dips in the SFR, which occur because it takes some time after we increase the resolution for gas to collapse past the new, higher star formation threshold, followed by sudden bursts as a large mass of gas simultaneously reaches the threshold; feedback then pushes the system back into equilibrium \citep[see also][]{FujimotoEtAl2018}. The total CR energy shows no comparable dip or burst at refinement, most likely because CRs are injected only through supernovae, so that changes in the SFR are imprinted on the CR energy only after a delay. In the high-resolution phase, the total CR energy depends on the CR diffusion coefficient: in Model~1 (high $\kappa_{\rm CR}$, identical to the value used in the low-resolution phase) the CR energy remains comparable to its pre-refinement level, increasing only mildly to just below ${\sim}5 \times 10^{55}$~erg by $t = 600$~Myr, whereas in Models~2 and~3, where $\kappa_{\rm CR}$ is reduced, CRs are confined more readily within the galactic disc and the total CR energy increases to just below ${\sim}1 \times 10^{56}$~erg (Model~2) and to somewhat above ${\sim}1 \times 10^{56}$~erg (Model~3). The mild increase seen even in Model~1 toward $t \simeq 600$~Myr is consistent with the delayed response of the CR energy to the preceding rise in the SFR. Continuing the simulations beyond $t=600$~Myr, we find that the rise in the total CR energy seen toward $t \simeq 600$~Myr does not persist: its growth slows markedly thereafter, and the total CR energies subsequently become approximately steady. We use the $t=600$~Myr snapshots as the reference epoch for the analysis below, corresponding to 100~Myr of evolution in the high-resolution phase. In contrast to the clear dependence of the CR energy on $\kappa_{\rm CR}$, the SFR histories of Models~1, 2, and 3 remain broadly similar throughout the high-resolution phase, with only a slightly lower SFR in Model~3 after $t \simeq 600$~Myr, as discussed further below.

\begin{figure*}
\centering
\includegraphics[width=0.97\textwidth]{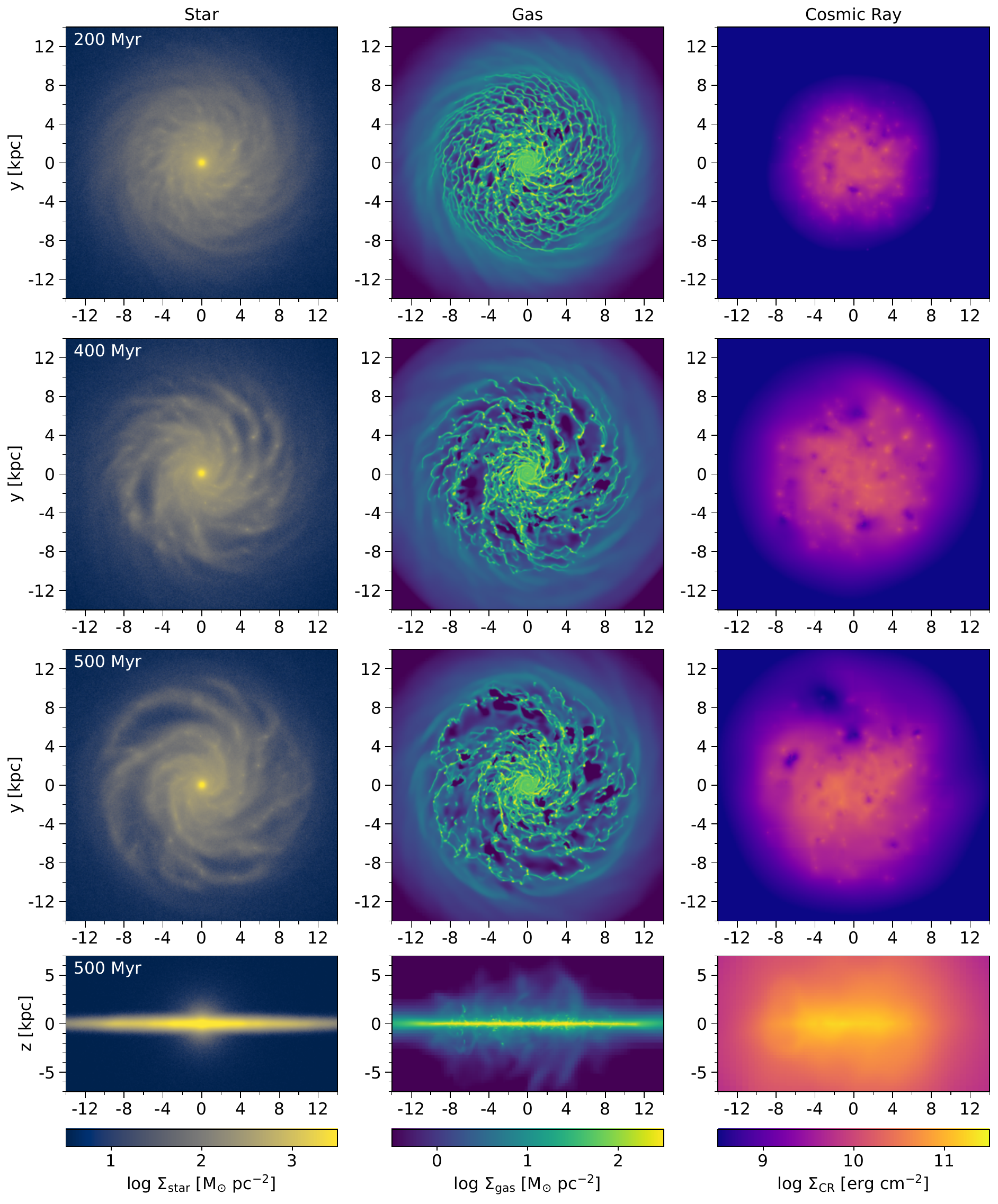}
\caption{
Morphology of the galactic disc during the low-resolution relaxation phase. From left to right, the panels show the stellar mass surface density, gas mass surface density, and CR energy surface density, each integrated over $-0.5 < z < 0.5$~kpc. From top to bottom, the disc is shown at $t = 200$, 400, and 500~Myr. The bottom row presents the disc viewed edge-on at 500~Myr. 
}
\label{fig:disk_evolution}
\end{figure*}

Figure~\ref{fig:disk_evolution} displays the face-on distributions of stars, gas, and CRs in the galactic disc at $t = 200$, 400, and 500~Myr, together with edge-on views at 500~Myr. The smooth, axisymmetric initial profiles (Section~\ref{subsec:initial_conditions}) develop mild non-axisymmetric features by $t = 200$~Myr, and by $t \sim 400$--$500$~Myr a multi-armed spiral pattern is fully developed in the stellar component and traced by the gas, which has formed a high-contrast, multi-phase ISM shaped by galactic shear, radiative cooling and heating, gas self-gravity, and stellar feedback. The CR surface density, initially concentrated toward the centre, becomes more radially extended as star formation spreads across the disc, with mild inhomogeneities at supernova injection sites. Both the stellar and gaseous morphologies and the global CR distribution change only modestly after $t \gtrsim 400$~Myr. The edge-on views show thin, rotationally supported stellar and gaseous discs and a more vertically extended CR distribution, confirming that the large-scale morphology evolves little once the quasi-steady state is reached.

\begin{figure*}
\centering
\includegraphics[width=1.0\textwidth]{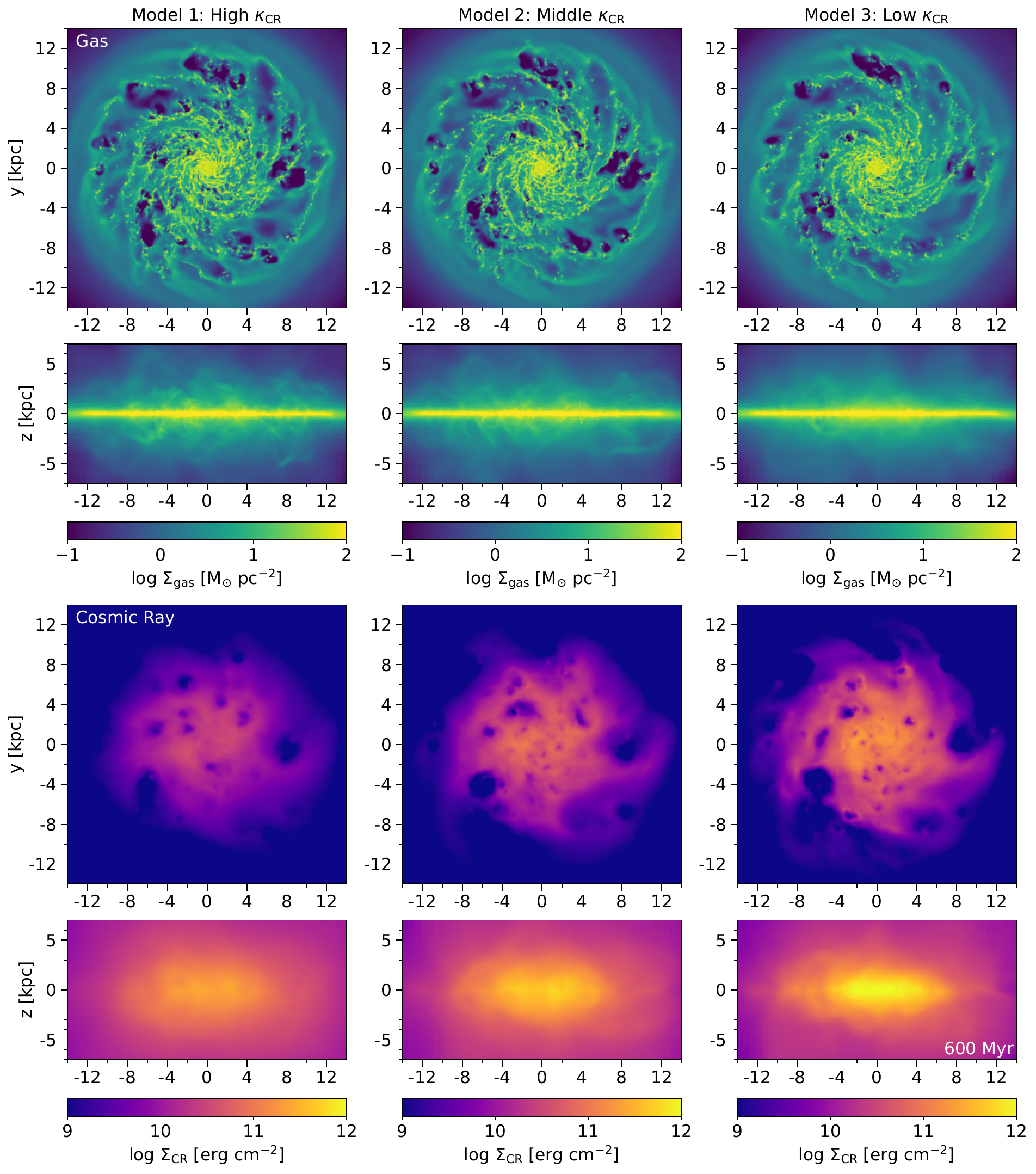}
\caption{
Morphology of the galactic disc during the high-resolution phase at $t = 600$~Myr. From left to right, the columns show Model~1 (High~$\kappa_{\rm CR}$), Model~2 (Middle~$\kappa_{\rm CR}$), and Model~3 (Low~$\kappa_{\rm CR}$). The top two rows show the gas mass surface density in the face-on (first row) and edge-on (second row) projections, and the bottom two rows show the CR energy surface density in the face-on (third row) and edge-on (fourth row) projections, integrated as in Figure~\ref{fig:disk_evolution}.
}
\label{fig:disk_model_comparison}
\end{figure*}

Figure~\ref{fig:disk_model_comparison} compares the gas and CR distributions in the high-resolution phase at $t = 600$~Myr among Models~1, 2, and 3, which differ in the CR diffusion coefficient $\kappa_{\rm CR}$. The most prominent difference between the models is in the CR distribution: as $\kappa_{\rm CR}$ decreases, CRs are confined more closely to the galactic mid-plane. This trend is evident in both the face-on and edge-on maps. In the face-on maps, the overall CR energy density is systematically enhanced from Model~1 to Model~3, while the radial extent of the distribution remains broadly similar among the models. The edge-on maps show that, as $\kappa_{\rm CR}$ decreases, the CR energy density near the mid-plane becomes substantially higher, accompanied by a reduction in the vertical extent of the CR distribution. This morphological dependence is fully consistent with the behaviour of the total CR energy shown in Figure~\ref{fig:time_evolution_SFR_CR}, where a lower $\kappa_{\rm CR}$ leads to a larger CR energy retained within the disc.

A close comparison of the gas maps also suggests that the large bubble-like cavities, presumably carved by stellar feedback, appear somewhat less prominent in Model~3 than in Models~1 and 2; the edge-on gas maps similarly show comparatively less gas lifted above the disc plane in Model~3. These differences may be associated with the enhanced CR pressure support and/or CR-driven gas removal that accompany a higher mid-plane CR energy density, which can act to regulate star formation and modulate the strength of the resulting stellar feedback \citep[e.g.,][]{SalemBryan2014}. Consistent with this picture, Figure~\ref{fig:time_evolution_SFR_CR} shows a comparatively lower SFR in Model~3 after $t \simeq 600$~Myr. We emphasize, however, that the differences in the gas morphology and SFR among the models are subtle, and that a detailed characterization of the interplay between CR transport, star formation, and feedback is beyond the scope of the present work, whose focus is the galactic-scale distribution of $^{10}$Be.

\subsection{\texorpdfstring{$^{10}$Be}{10Be} distribution in the galactic disc} 
\label{subsec:10Be_distribution_in_the_galactic_disc}

\begin{figure*}
\centering
\includegraphics[width=\textwidth]{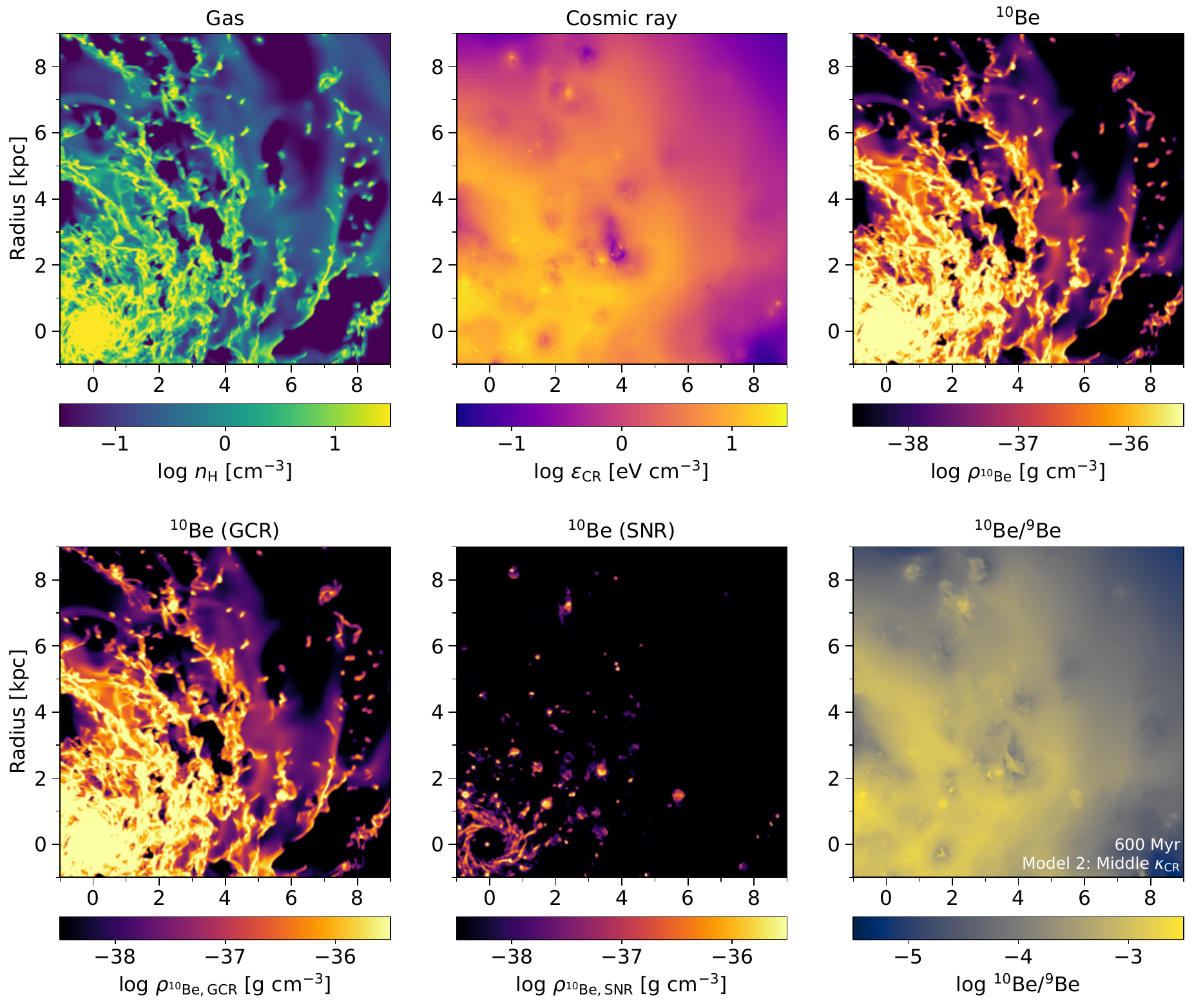}
\caption{
Zoom-in views of the galactic disc at $t = 600$~Myr for Model~2, the intermediate-diffusivity case. The panels show the gas density (top left), CR energy density (top middle), $^{10}$Be density (top right), GCR-produced $^{10}$Be density (bottom left), SNR-produced $^{10}$Be density (bottom middle), and the $^{10}$Be/$^{9}$Be abundance ratio (bottom right). All quantities are projected along the line of sight and weighted by the gas density, integrated over $-200 < z < 200$ pc. The galactic centre is located at the lower-left corner at $(x, y) = (0, 0)$ kpc, and the galactic disc rotates clockwise.
}
\label{fig:zoom_projections}
\end{figure*}

\begin{figure*}
\centering
\includegraphics[width=0.48\textwidth]{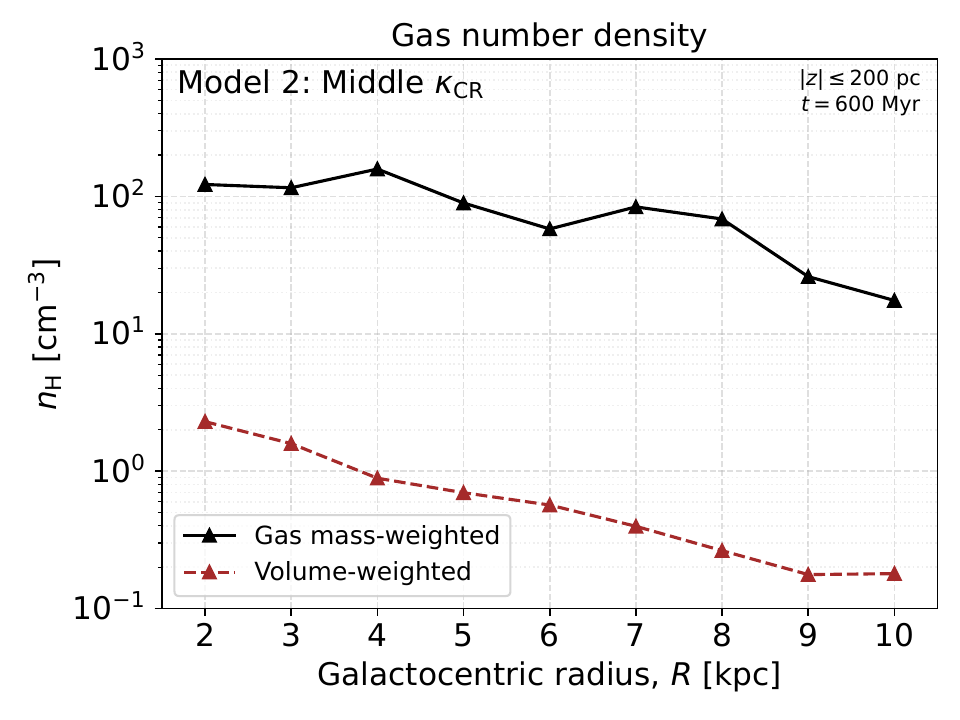}
\includegraphics[width=0.48\textwidth]{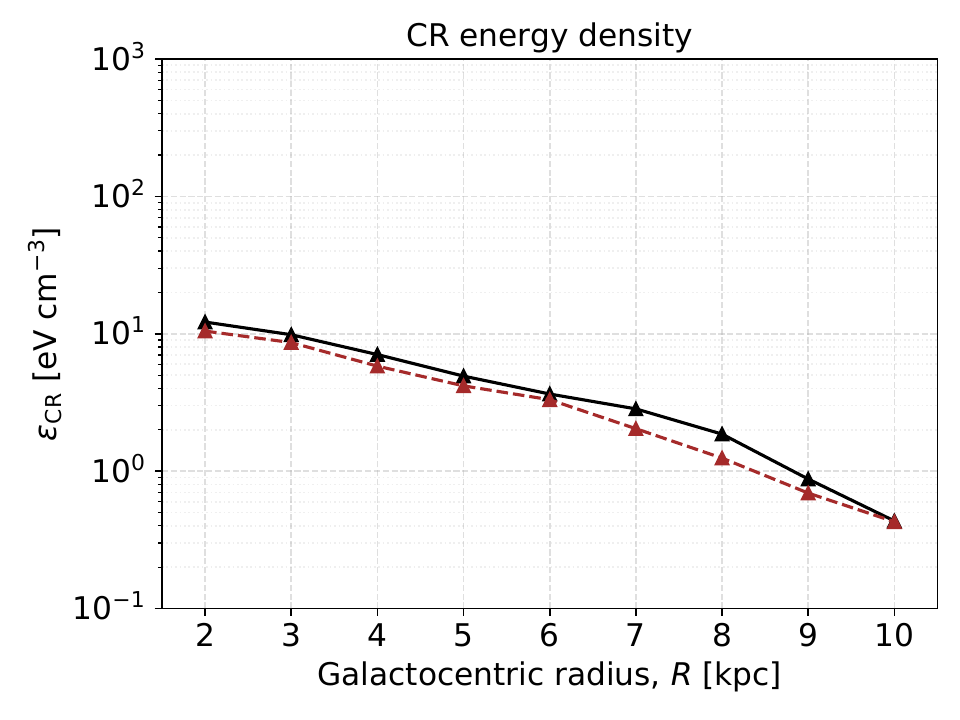}
\includegraphics[width=0.48\textwidth]{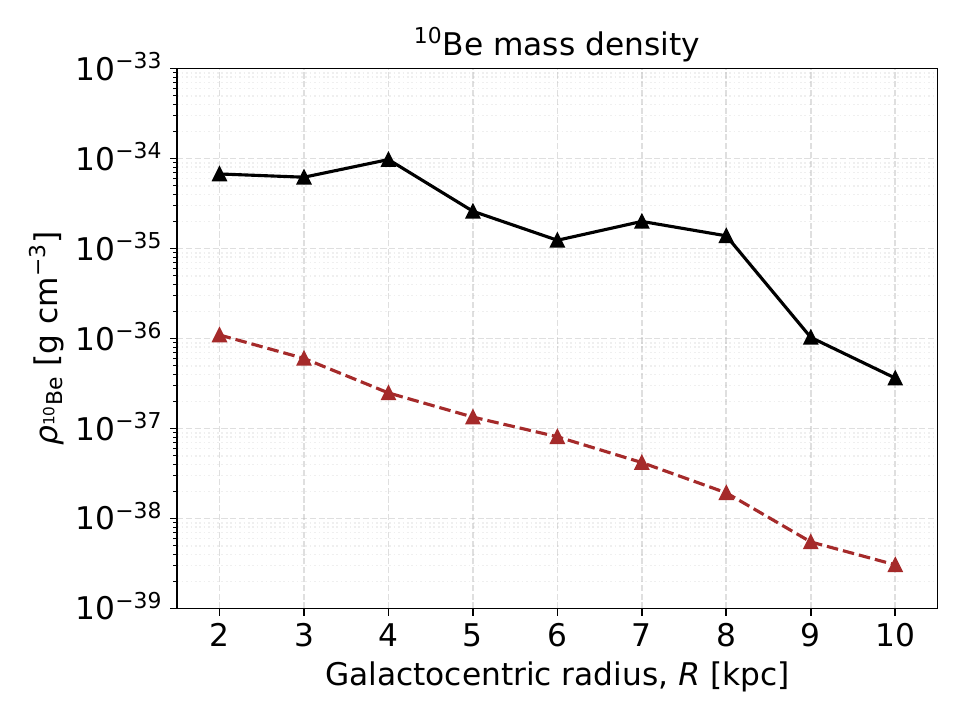}
\includegraphics[width=0.48\textwidth]{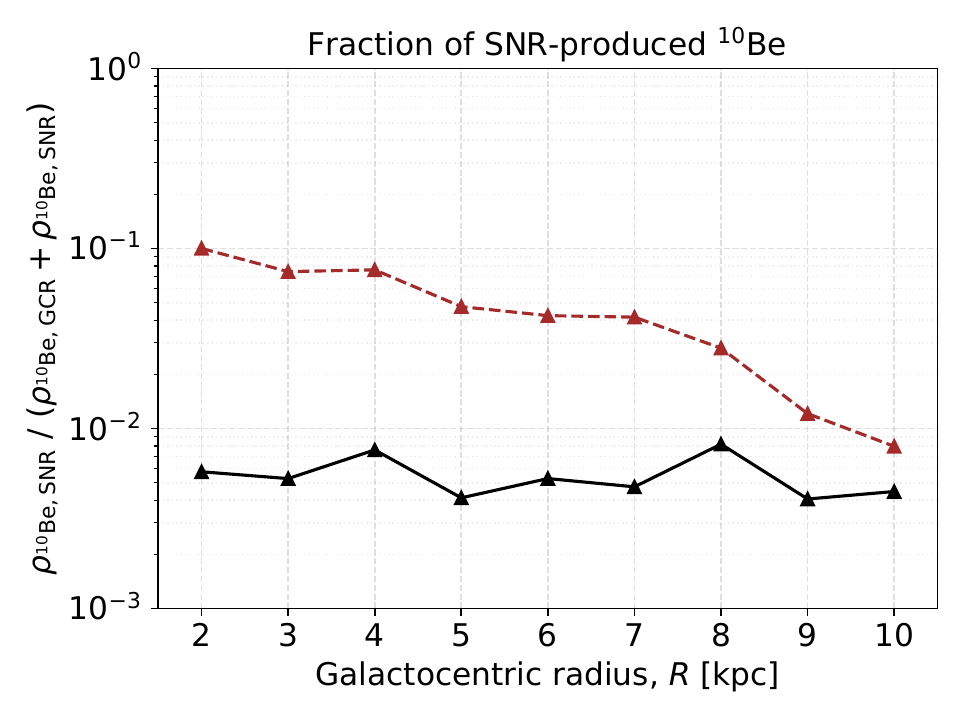}
\caption{
Azimuthally averaged radial profiles of the galactic disc at $t = 600$~Myr for Model~2, the intermediate-diffusivity case. The four panels show the gas number density (top left), CR energy density (top right), $^{10}$Be mass density (bottom left), and the fraction of SNR-produced $^{10}$Be in the total $^{10}$Be (bottom right). Each quantity is shown as a weighted mean, either weighted by the gas mass of each cell (black solid lines) or by the cell volume (brown dashed lines). The radial bin size is 1~kpc, and only cells within $|z| \le 200$~pc are included.
}
\label{fig:radial_profile_model_2}
\end{figure*}

To first understand the basic characteristics of the $^{10}$Be distribution in the galactic disc, for illustration we show Model~2, the intermediate-diffusivity case, at $t = 600$~Myr in this section and discuss its spatial and radial properties.
The spatial distribution of $^{10}$Be shows a gas-like local morphology with a CR-like radial trend.  
In the zoom-in projection maps (Figure~\ref{fig:zoom_projections}), the $^{10}$Be density exhibits filamentary, high-contrast structures that closely resemble the gas morphology rather than the relatively smooth CR distribution.  
This gas-like spatial contrast is quantitatively confirmed by the radial profile of the $^{10}$Be density (bottom-left panel of Figure~\ref{fig:radial_profile_model_2});  
we characterize radial distributions using two types of weighted means: a gas mass-weighted mean, which weights each grid cell by its gas mass and thus emphasizes dense, star-forming regions, and a volume-weighted mean, which weights each cell equally by volume and thus better reflects the diffuse component of the ISM.
The panel shows that the gas mass-weighted mean of the $^{10}$Be density is more than two orders of magnitude higher than the volume-weighted mean across most radii, mirroring the large separation between the mass- and volume-weighted gas densities (top-left panel of Figure~\ref{fig:radial_profile_model_2}).  
These trends indicate that $^{10}$Be is concentrated in compact, dense environments such as molecular clouds and star-forming regions, where the gas dominates the mass-weighted average.

Despite this resemblance to the gas in terms of spatial contrast, the radial decline of $^{10}$Be is much steeper than that of the gas.  
Whereas both the mass- and volume-weighted gas densities vary only mildly---by about one order of magnitude or less---between $R = 2$ and $10\ \mathrm{kpc}$, the $^{10}$Be mass density decreases by more than two orders of magnitude over the same range. 

The top-right panel of Figure~\ref{fig:radial_profile_model_2} shows that the CR energy density declines by about 1.5 orders of magnitude between $R = 2$ and $10$~kpc. At the Solar radius ($R = 8$~kpc), the CR energy density is $\sim 1$--$2$~eV~cm$^{-3}$ in Model~2, consistent with the locally observed value of $\sim 1$~eV~cm$^{-3}$ \citep{Webber1998, CummingsEtAl2016}, indicating that Model~2 reproduces a realistic CR energy density in the Solar neighbourhood. 
The radial gradient of CR energy density broadly resembles the steep gradient of $^{10}$Be, although it is itself less steep than the more than two orders of magnitude decline of $^{10}$Be, suggesting that the large-scale behaviour of $^{10}$Be reflects the combined radial decline of both the gas and CR distributions, rather than being set by either alone.

To clarify this connection, we examine the relative contributions from the two production channels of $^{10}$Be: (i) GCR-produced $^{10}$Be, originating from interactions between CRs and the ISM, and (ii) SNR-produced $^{10}$Be, injected directly by supernova remnants. The zoom-in maps (Figure~\ref{fig:zoom_projections}) show that the GCR-produced component overwhelmingly dominates the total $^{10}$Be distribution. The radial profile of the SNR-produced fraction (bottom-right panel of Figure~\ref{fig:radial_profile_model_2}) confirms this quantitatively: SNR-produced $^{10}$Be accounts for up to about ten percent of the total in the volume-weighted mean, and remains below one percent at all radii in the gas mass-weighted mean that emphasizes the high-density, star-forming regions. This weighting dependence likely arises because the SNR-produced $^{10}$Be is released into the surrounding ISM as the remnant expands (Section~\ref{subsec:10Be_production_model}) and is thereby spread into the low-density gas, whereas the GCR-produced $^{10}$Be closely follows the dense gas in which it is predominantly produced. Because the GCR production rate of $^{10}$Be is proportional to the product of the gas density and the CR energy density (Equation~\ref{eq:rho10Be_evolution}), the resulting $^{10}$Be distribution inherits the high-contrast, clumpy morphology of the gas while following a radial decline steeper than that of either field alone.

\subsection{Model dependence of \texorpdfstring{$^{10}$Be}{10Be} distribution} 
\label{subsec:model_dependence_10Be_distribution}

\begin{figure*}
\centering
\includegraphics[width=0.48\textwidth]{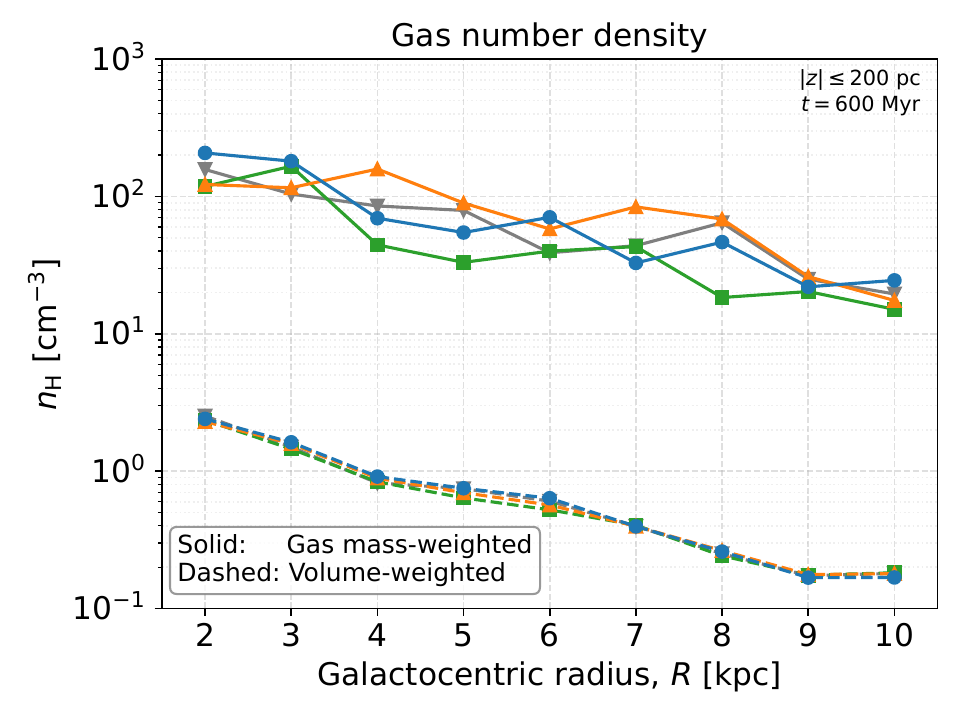}
\includegraphics[width=0.48\textwidth]{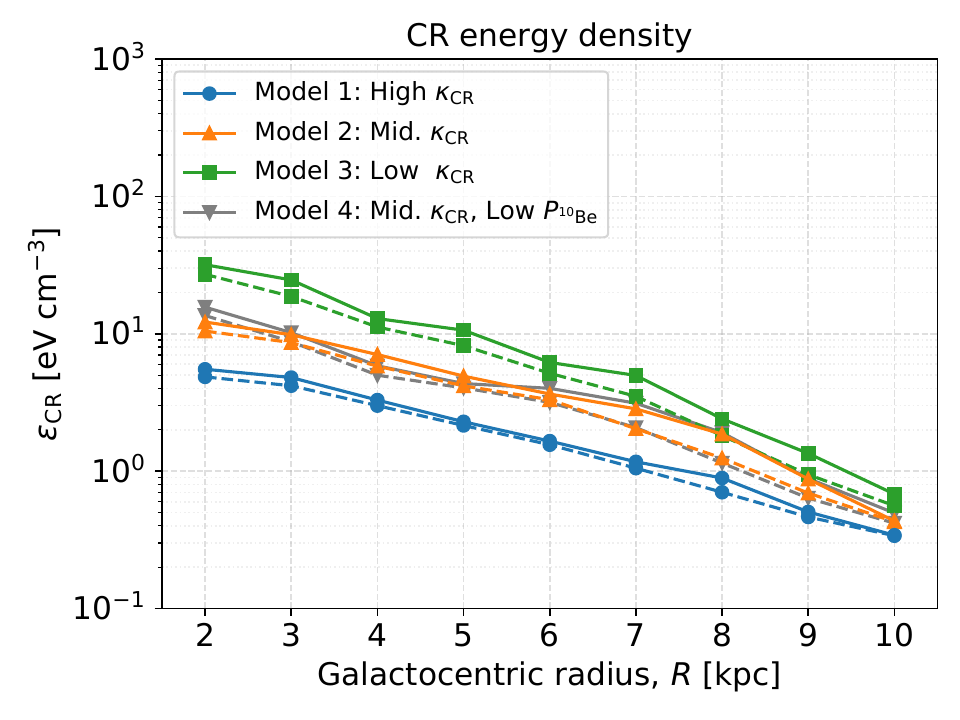}
\includegraphics[width=0.48\textwidth]{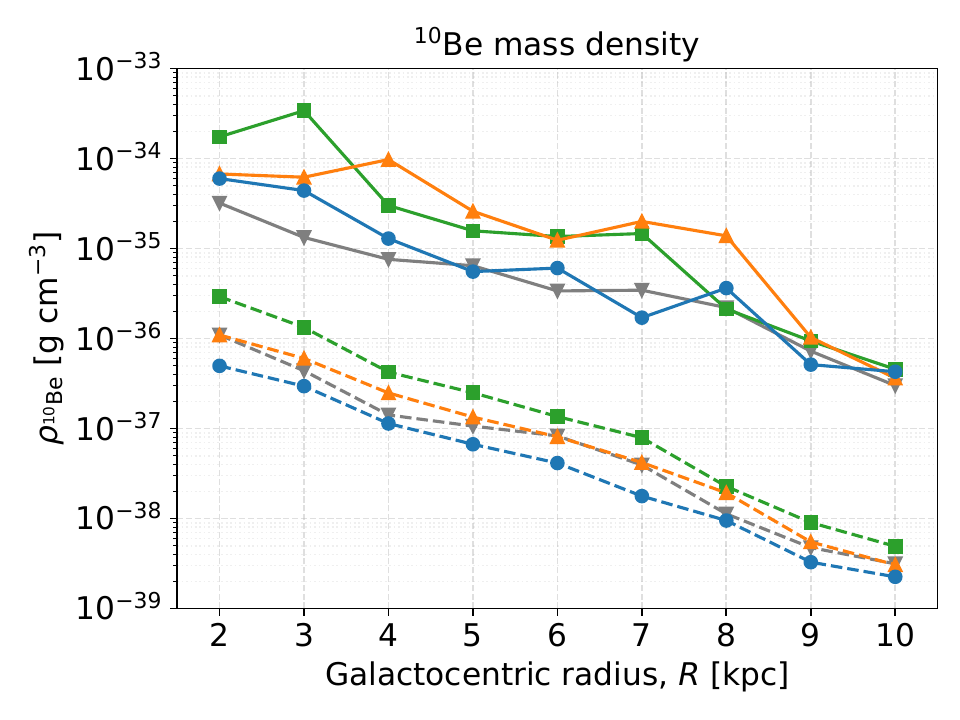}
\includegraphics[width=0.48\textwidth]{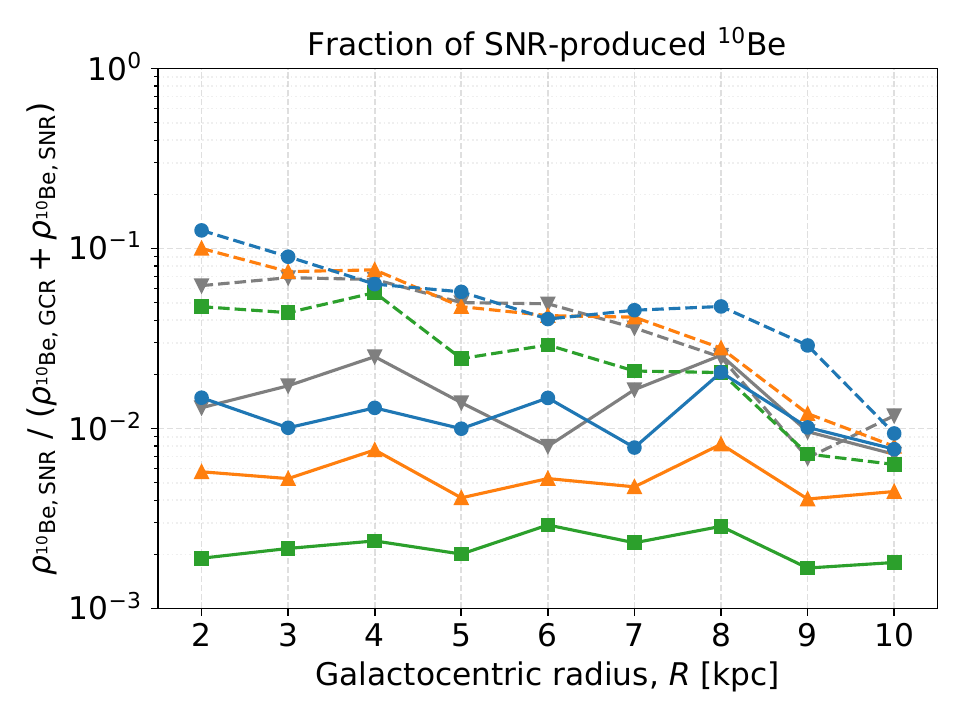}
\caption{
Same as Figure~\ref{fig:radial_profile_model_2}, but comparing the four models at $t = 600$~Myr. The curves represent Model~1 (High~$\kappa_{\rm CR}$; blue circles), Model~2 (Middle~$\kappa_{\rm CR}$; orange triangles), Model~3 (Low~$\kappa_{\rm CR}$; green squares), and Model~4 (Middle~$\kappa_{\rm CR}$, with a reduced GCR production rate of $^{10}$Be in dense gas; grey inverted triangles). For each model, the gas mass-weighted mean is shown by solid lines and the volume-weighted mean by dashed lines. 
}
\label{fig:radial_profile_model_comparison}
\end{figure*}

Here we examine how the $^{10}$Be distribution depends on the model parameters by comparing the azimuthally averaged radial profiles among the four models (Figure~\ref{fig:radial_profile_model_comparison}). Models~1, 2, and 3 differ in the CR diffusion coefficient $\kappa_{\rm CR}$, while Model~4 shares the same $\kappa_{\rm CR}$ as Model~2 and differs only in that the GCR production rate of $^{10}$Be is reduced by a factor of ten in dense gas ($n_{\rm H} > 100\ \mathrm{cm^{-3}}$).

The gas number density shows no clear systematic dependence on $\kappa_{\rm CR}$. In the volume-weighted profile, the differences among the models are negligible. In the gas mass-weighted profile, small differences appear between the models at individual radii, but within each model the profile also exhibits radius-to-radius scatter that is comparable in magnitude to the differences between the models. This is illustrated by the fact that Models~2 and 4, which share the same gas dynamics and are therefore expected to yield identical gas distributions, do not coincide exactly at all radii in the gas mass-weighted profile. We attribute this to fluctuations arising from the stochastic nature of the star formation and feedback prescriptions. Because the difference between Models~2 and 4 is comparable to the differences between the other models, we conclude that the gas distribution is largely insensitive to $\kappa_{\rm CR}$. The only systematic trend is that Model~3 exhibits a slightly lower gas density than the other models at most radii, which is likely related to its slightly lower SFR after $t \simeq 600$~Myr seen in Figure~\ref{fig:time_evolution_SFR_CR}.

In contrast, the CR energy density shows a clear systematic dependence on $\kappa_{\rm CR}$: it increases as $\kappa_{\rm CR}$ decreases, in the order Model~3 $>$ Model~2 $>$ Model~1. The radial decline is also slightly steeper for lower $\kappa_{\rm CR}$, although the difference is barely discernible by eye: linear fits to $\log_{10}\epsilon_{\rm CR}$ over $R = 2$--$10$~kpc give slopes of $\simeq -0.15$, $-0.17$, and $-0.21$~dex~kpc$^{-1}$ for Models~1, 2, and 3, respectively. CRs with a lower diffusivity remain more closely confined to their injection sites, which are concentrated in the inner, star-forming disc, whereas a higher diffusivity smooths out the radial distribution. Model~4 nearly coincides with Model~2, as expected, since the two models share the same $\kappa_{\rm CR}$ and the modified $^{10}$Be production rate has no effect on the CR field itself. This ordering is consistent with the model dependence of the total CR energy within the disc shown in Figure~\ref{fig:time_evolution_SFR_CR}. Despite this systematic dependence, all four models yield CR energy densities at the Solar radius ($R = 8$~kpc) of order $\sim 1$~eV~cm$^{-3}$, consistent with the locally observed value \citep{Webber1998, CummingsEtAl2016}.

The $^{10}$Be mass density reflects the combined behaviour of the gas and CR distributions, and its dependence on $\kappa_{\rm CR}$ differs between the two weighting schemes. In the gas mass-weighted profile, the difference between Models~1 and 2 indicates that a lower $\kappa_{\rm CR}$ tends to produce a higher $^{10}$Be density, following the higher CR energy density. Model~3, however, does not show a systematically higher $^{10}$Be mass density than Model~2; its slightly lower gas density in the gas mass-weighted profile appears to offset the effect of its higher CR energy density, so that the ordering among the three models is no longer monotonic. In the volume-weighted profile, by contrast, the differences among the models are smaller but follow a clean ordering, Model~3 $>$ Model~2 $>$ Model~1. This is a natural consequence of the production physics, in which the GCR-produced $^{10}$Be is proportional to the product of the gas density and the CR energy density: in the volume-weighted profile the gas number density shows almost no difference among the models, so that the model dependence of the $^{10}$Be density essentially reflects that of the CR energy density, which increases in the order Model~3 $>$ Model~2 $>$ Model~1. 

The effect of the reduced GCR production rate in dense gas is seen by comparing Models~2 and 4. In the gas mass-weighted profile, Model~4 shows a lower $^{10}$Be mass density than Model~2, as expected. The reduction, however, appears to be less than a factor of ten overall, even though the production rate was reduced by a factor of ten in the dense gas. This may be because the SNR-produced $^{10}$Be, whose production rate is unchanged between Models~2 and 4, could partially compensate for the reduction; in addition, GCR-produced $^{10}$Be in gas below the density threshold ($n_{\rm H} = 100\ \mathrm{cm^{-3}}$), where the production rate is unchanged, may also contribute to the gas mass-weighted average. The resulting $^{10}$Be mass density of Model~4 is comparable to that of Model~1.
In the volume-weighted profile, by contrast, the difference between Models~2 and 4 is negligible: because $^{10}$Be production is suppressed only in the dense gas ($n_{\rm H} > 100~\mathrm{cm^{-3}}$), while the volume-weighted average is dominated by the more voluminous low-density gas---where the production rate is unchanged from Model~2---the suppression has little effect on the volume-weighted $^{10}$Be mass density.

Finally, the fraction of SNR-produced $^{10}$Be differs among the models, with Model~3 showing the lowest values in the gas mass-weighted profile. This is a natural consequence of the production physics: the SNR-produced $^{10}$Be does not depend on the CR distribution and is governed only by the SFR, whereas the GCR-produced $^{10}$Be scales with the CR energy density. Because Model~3 has the highest CR energy density owing to its low $\kappa_{\rm CR}$, the relative contribution of the SNR-produced component is correspondingly the smallest.

\subsection{\texorpdfstring{$^{10}$Be/$^{9}$Be}{10Be/9Be} distribution} 
\label{subsec:10Be_9Be_distribution}

\begin{figure*}
\centering
\includegraphics[width=0.48\textwidth]{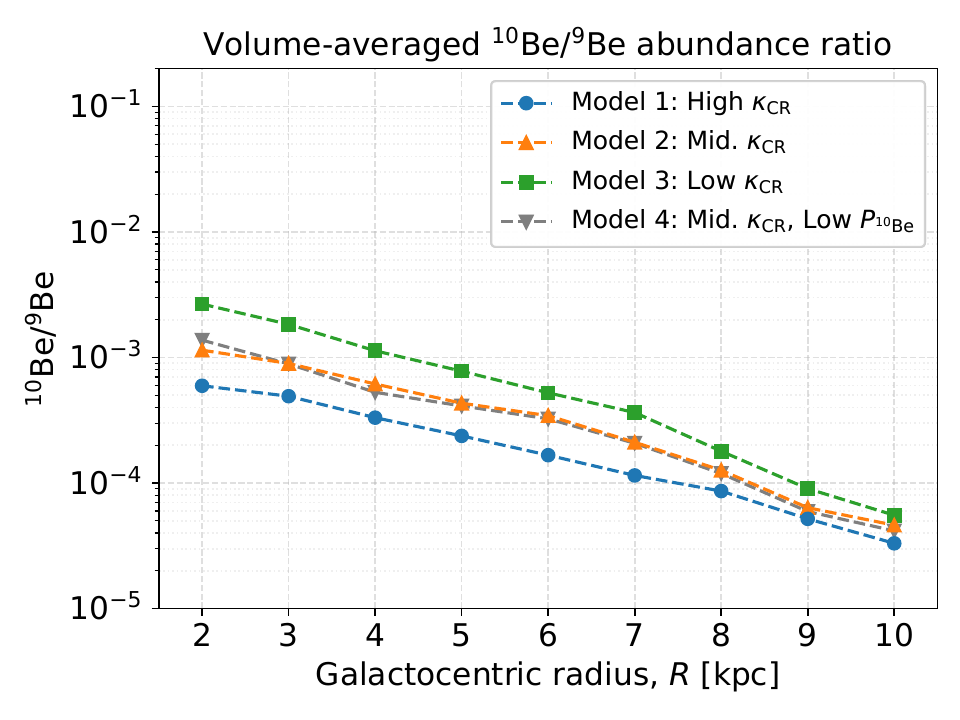}
\includegraphics[width=0.48\textwidth]{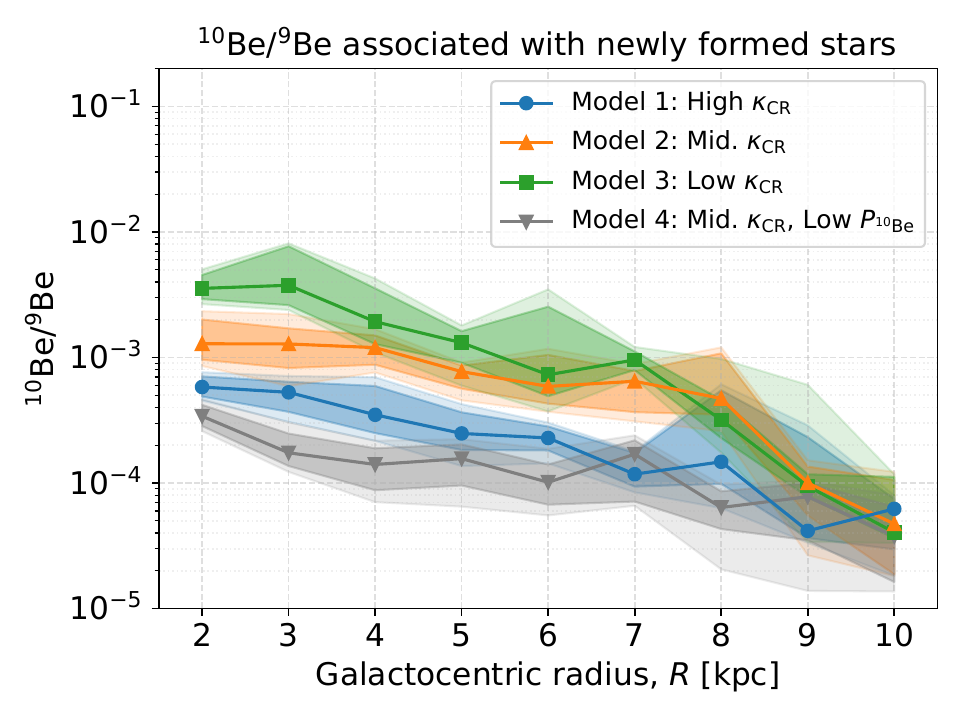}
\caption{
Radial profiles of the $^{10}$Be/$^{9}$Be abundance ratio at $t = 600$~Myr for the four models: Model~1 (High~$\kappa_{\rm CR}$; blue circles), Model~2 (Middle~$\kappa_{\rm CR}$; orange triangles), Model~3 (Low~$\kappa_{\rm CR}$; green squares), and Model~4 (Middle~$\kappa_{\rm CR}$, with a reduced GCR production rate of $^{10}$Be in dense gas; grey inverted triangles). \textit{Left}: the volume-weighted mean $^{10}$Be/$^{9}$Be ratio of the gas, azimuthally averaged in radial bins of 1~kpc with only cells within $|z| \le 200$~pc included (see the footnote in Section~\ref{subsec:10Be_9Be_distribution} for why only the volume-weighted mean is shown). \textit{Right}: the $^{10}$Be/$^{9}$Be ratio inherited by newly formed stars (formed within the last 5~Myr); the solid lines show the median in each radial bin, and the darker and lighter shaded regions show the 25th--75th and 16th--84th percentile ranges, respectively. In both panels, the $^{9}$Be abundance is assumed to be spatially uniform at the Solar photospheric value.
The meteoritic value is not indicated here because the simulated ratios refer to the present-day Galaxy; the comparison with the meteoritic value, which requires the correction for the higher star formation rate at the epoch of Solar System formation, is made in Figure~\ref{fig:initial_10Be_9Be_solar_birth_radius}.
}
\label{fig:radial_profile_10Be_9Be}
\end{figure*}

In this section, we present the $^{10}$Be/$^{9}$Be ratio, which can be compared directly with the meteoritic value inferred for the early Solar System. As described in Section~\ref{subsubsec:9Be_treatment}, our simulation does not solve for the production and transport of the stable isotope $^{9}$Be, so the $^{10}$Be/$^{9}$Be ratio is obtained in post-processing by combining the simulated $^{10}$Be density with an assumed $^{9}$Be abundance. Throughout this section we adopt the simplest assumption, namely that the $^{9}$Be abundance is spatially uniform across the disc, fixed to the Solar photospheric value, $^{9}$Be/H $= 2.4\times10^{-11}$ \citep{AsplundEtAl2009}, which we use as a reference abundance. The comparison with the meteoritic value, including the enhanced $^{10}$Be production associated with a possibly higher star formation rate at the epoch of Solar System formation 4.6~Gyr ago, is presented in Section~\ref{subsec:initial_10Be_9Be_solar_birth_environment}, and the effect of a possible galactocentric gradient of $^{9}$Be is examined in Section~\ref{subsec:inherited_10Be_9Be_across_disc}.

Unlike the total $^{10}$Be density, which---being overwhelmingly GCR-produced (Section~\ref{subsec:10Be_distribution_in_the_galactic_disc})---inherits spatial characteristics from both the gas and CR components, the distribution of the $^{10}$Be/$^{9}$Be abundance ratio is largely governed by the CR field.  
In the zoom-in projection maps (Figure~\ref{fig:zoom_projections}), the $^{10}$Be/$^{9}$Be ratio shows none of the high-contrast, filamentary substructure characteristic of the gas or of the $^{10}$Be density.  
Instead, it exhibits a smooth morphology with a clear radial decline, closely resembling the spatial appearance of the CR energy density.  

This behaviour is quantitatively confirmed by the radial profile of the $^{10}$Be/$^{9}$Be ratio (left panel of Figure~\ref{fig:radial_profile_10Be_9Be})\footnote{For the $^{10}$Be/$^{9}$Be ratio of the gas we show only the volume-weighted mean. Because $^{10}$Be/$^{9}$Be is, under the assumption of a spatially uniform $^{9}$Be/H, proportional to the per-cell ratio of the $^{10}$Be density to the gas density, its gas mass-weighted mean reduces to the total $^{10}$Be divided by the total hydrogen within each radial bin, and therefore no longer emphasizes the dense gas as it does for the other quantities.}. For Model~2, the volume-weighted mean $^{10}$Be/$^{9}$Be ratio declines smoothly with radius, decreasing by about 1.5 orders of magnitude between $R = 2$ and $10\ \mathrm{kpc}$. This decline closely follows that of the CR energy density (top-right panel of Figure~\ref{fig:radial_profile_model_comparison}), which decreases by a comparable amount over the same range, rather than the much steeper decline of the $^{10}$Be mass density, which decreases by about 2.5 orders of magnitude. This smooth radial behaviour, together with the smooth spatial morphology seen in the zoom-in maps, confirms that the $^{10}$Be/$^{9}$Be ratio is CR-like.

This correspondence follows from the production physics. Because nearly all of the $^{10}$Be in the disc is GCR-produced (Section~\ref{subsec:10Be_distribution_in_the_galactic_disc}), at a rate proportional to the product of the gas density and the CR energy density, the $^{10}$Be/$^{9}$Be ratio---obtained by normalizing the $^{10}$Be density by the hydrogen density under the assumption of a uniform $^{9}$Be/H---depends primarily on the CR energy density, and thus inherits the smooth morphology and radial decline of the CR field rather than the clumpy, multi-phase structure of the gas.

We now examine the model dependence of the $^{10}$Be/$^{9}$Be ratio. At all radii, the $^{10}$Be/$^{9}$Be ratio increases as $\kappa_{\rm CR}$ decreases, in the order Model~3 $>$ Model~2 $>$ Model~1, mirroring the model dependence of the CR energy density (top-right panel of Figure~\ref{fig:radial_profile_model_comparison}). This is consistent with the $^{10}$Be/$^{9}$Be ratio being governed primarily by the CR field. Model~4, in which the GCR production rate of $^{10}$Be is reduced by a factor of ten in dense gas, shows a $^{10}$Be/$^{9}$Be ratio that is nearly indistinguishable from that of Model~2 at most radii, despite the reduced production rate. This may seem counterintuitive, but it can be understood as a consequence of the volume weighting, for the same reason discussed in Section~\ref{subsec:model_dependence_10Be_distribution} for the $^{10}$Be mass density: the suppression, being confined to the dense gas, has little effect on the volume-weighted average, which is dominated by the low-density gas where the production rate is unchanged. As a result, the volume-weighted $^{10}$Be/$^{9}$Be ratios of Models~2 and 4 are nearly identical.

So far we have considered the volume-weighted mean $^{10}$Be/$^{9}$Be ratio of the gas. We now turn to the $^{10}$Be/$^{9}$Be ratio inherited by newly formed stars (right panel of Figure~\ref{fig:radial_profile_10Be_9Be}), where each star is assigned the $^{10}$Be/$^{9}$Be value of the gas cell in which it formed; throughout this paper, ``newly formed stars'' denotes star particles formed within the last 5~Myr before the snapshot at $t = 600$~Myr. Because stars form in dense gas, this stellar sample preferentially traces it; in this respect it plays a role similar to the gas mass-weighted mean discussed in Sections~\ref{subsec:10Be_distribution_in_the_galactic_disc} and \ref{subsec:model_dependence_10Be_distribution}, while tracing the actively star-forming gas more directly. As with the CR energy density profile (top-right panel of Figure~\ref{fig:radial_profile_model_comparison}), the difference between the volume-weighted mean of the gas (left panel of Figure~\ref{fig:radial_profile_10Be_9Be}) and the stellar values (right panel) is modest, and the median stellar $^{10}$Be/$^{9}$Be ratio declines by somewhat more than an order of magnitude between $R = 2$ and $10\ \mathrm{kpc}$, mirroring the radial trend of the volume-weighted mean. A closer look reveals that, particularly for Models~2 and 3, the median stellar ratio is systematically higher than the volume-weighted mean of the gas at the same radius---by a factor of $\sim 5$ at the Solar radius for Model~2. This likely reflects the fact that newly formed stars trace the dense, star-forming regions directly, where the local CR energy density---and hence the $^{10}$Be/$^{9}$Be ratio---tends to be enhanced. For the same reason, the suppression of $^{10}$Be production in dense gas in Model~4 is more clearly imprinted on the stellar ratio than on the volume-weighted mean of the gas: unlike in the left panel, the median stellar $^{10}$Be/$^{9}$Be ratio of Model~4 is noticeably lower than that of Model~2. 

A detailed comparison with the meteoritic value, and its implications for the Solar birth radius, are presented in Section~\ref{subsec:initial_10Be_9Be_solar_birth_environment}.

\section{Discussion} 
\label{sec:discussion}

\subsection{Initial \texorpdfstring{$^{10}$Be/$^{9}$Be}{10Be/9Be} ratio and the Solar birth environment}
\label{subsec:initial_10Be_9Be_solar_birth_environment} 

We now compare the $^{10}$Be/$^{9}$Be ratios inherited by newly formed stars with the meteoritic value (i.e., the initial $^{10}$Be/$^{9}$Be of the early Solar System recorded by meteoritic inclusions), and discuss the implications for the environment in which the Sun formed. We frame this comparison as a hypothesis test over the assumed Solar birth radius, $R_{\odot,\rm init}$: if the Sun formed at a galactocentric radius $R_{\odot,\rm init}$, is the $^{10}$Be/$^{9}$Be ratio inherited by stars forming there consistent with the meteoritic value?

For this test, the appropriate normalization of the stable isotope follows directly from the measurement itself: the meteoritic $^{10}$Be/$^{9}$Be is measured relative to the $^{9}$Be contained in the proto-Solar material, and the Solar photospheric abundance, $^{9}$Be/H $= 2.4\times10^{-11}$ \citep{AsplundEtAl2009}, samples precisely that material\footnote{Applying the standard protosolar correction of $+0.04$~dex for gravitational settling would give $^{9}$Be/H $\simeq 2.6\times10^{-11}$, lowering all $^{10}$Be/$^{9}$Be ratios by $\simeq 8$\%. This is small compared with the uncertainty in the Be abundance measurement ($\pm 0.09$~dex) and with the uncertainty in the CR-flux correction discussed below, and does not affect our conclusions.}. We therefore evaluate the inherited ratio at each assumed birth radius using this uniform Solar value, as in Section~\ref{subsec:10Be_9Be_distribution}. A radial gradient of $^{9}$Be/H, however plausible, does not enter this test, because anchoring the $^{9}$Be abundance to the protosolar value at the assumed birth radius makes any gradient cancel at the evaluation point; it matters instead for the ratios inherited elsewhere in the disc, which we examine in Section~\ref{subsec:inherited_10Be_9Be_across_disc}.

The $^{10}$Be side, by contrast, does require a correction, because our simulations reproduce the present-day Milky Way whereas the meteoritic value pertains to the epoch of Solar System formation 4.6~Gyr ago. Reconstructions of the Milky Way's star formation history do not uniquely determine the magnitude of the difference between the star formation rate at that epoch and the present-day value, owing to differences in time resolution, spatial coverage, and methodology. Some studies suggest that the star formation rate remained roughly constant after the onset of thin-disc formation \citep[e.g.,][]{HaywoodEtAl2016}, whereas several \textit{Gaia}-based studies indicate enhanced star formation around the epoch of Solar System formation \citep[e.g.,][]{MorEtAl2019,RuizLaraEtAl2020,AlzateEtAl2021}. A higher star formation rate implies a higher supernova rate, a higher CR production rate, and hence a higher CR energy density, which in turn raises the GCR production of $^{10}$Be. Assuming, as a first-order approximation for this sensitivity test, that the CR energy density scales linearly with the star formation rate, we multiply the $^{10}$Be of each star particle by an SFR enhancement factor $\Psi$, defined as the ratio of the star formation rate at the epoch of Solar System formation to the present-day value, and adopt $\Psi=2.0$ as a representative value. For example, the exponentially declining star formation history inferred by \citet{Bovy2017} implies an enhancement of $\simeq 1.9$ at a lookback time of 4.6~Gyr. Because the inherited $^{10}$Be/$^{9}$Be ratio scales linearly with $\Psi$, the results can be straightforwardly rescaled for alternative choices.

\begin{figure}
\centering
\includegraphics[width=0.98\columnwidth]{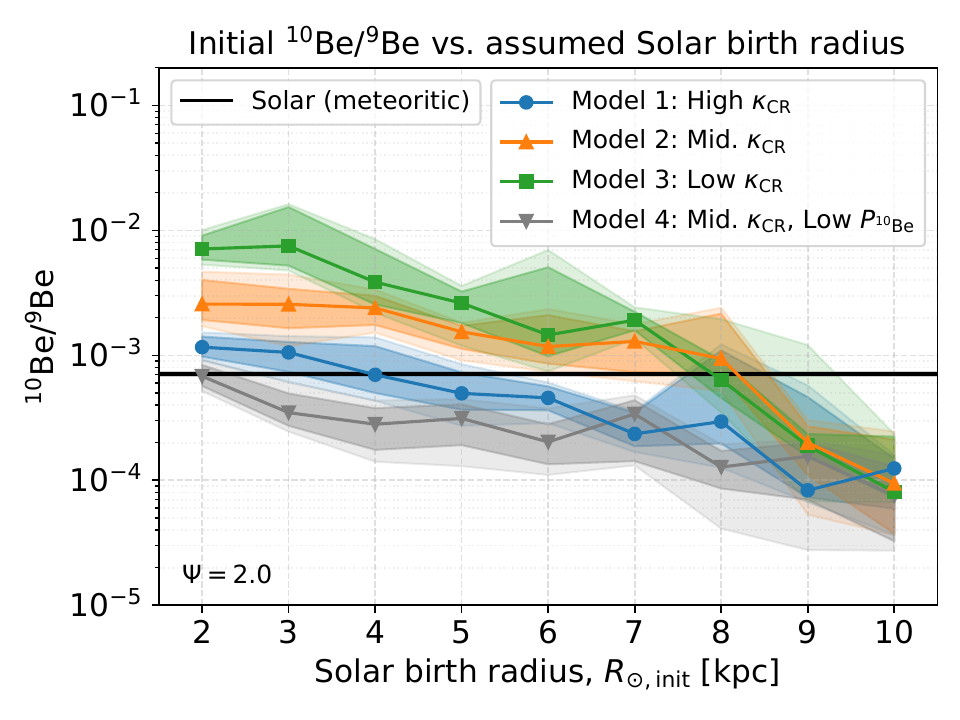}
\caption{
Initial $^{10}$Be/$^{9}$Be ratio inherited by newly formed stars, as a function of the assumed Solar birth radius $R_{\odot,\rm init}$, for the four models at $t = 600$~Myr. For each $R_{\odot,\rm init}$, the ratio is computed with a spatially uniform $^{9}$Be/H $= 2.4\times10^{-11}$ \citep{AsplundEtAl2009}, appropriate for the proto-Solar material, and the $^{10}$Be of each star particle is multiplied by an SFR enhancement factor $\Psi = 2.0$ to account for a possibly higher CR energy density at the epoch of Solar System formation (see text). Symbols show the median over stars formed within the last 5~Myr before $t = 600$~Myr in radial bins of 1~kpc, and the darker and lighter shaded regions show the 25th--75th and 16th--84th percentile ranges, respectively. The horizontal black line marks the initial Solar System value of $\sim 7\times10^{-4}$ inferred from meteoritic inclusions \citep{DunhamEtAl2022}. Note that this figure represents a scan over the hypothesis ``the Sun formed at $R_{\odot,\rm init}$,'' not a radial profile of the Galactic disc at that epoch (cf. Figure~\ref{fig:radial_profile_10Be_9Be_solar_epoch}).
}
\label{fig:initial_10Be_9Be_solar_birth_radius}
\end{figure}

Figure~\ref{fig:initial_10Be_9Be_solar_birth_radius} shows the resulting initial $^{10}$Be/$^{9}$Be as a function of the assumed Solar birth radius. We regard a model as consistent with the meteoritic value at a given birth radius when the meteoritic value falls within the 25th--75th percentile range of the stellar distribution. By this criterion, Model~1 is consistent with the meteoritic value at $R_{\odot,\rm init} \simeq 3$--$5$~kpc, and also at $R_{\odot,\rm init} = 8$~kpc, where the median lies below the meteoritic value but the broad star-to-star scatter brings the percentile range up to it; at $R_{\odot,\rm init} = 2$~kpc its distribution lies above the meteoritic value, and at $R_{\odot,\rm init} = 6$--$7$~kpc and $\gtrsim 9$~kpc it falls below. Model~2 is consistent with the meteoritic value at $R_{\odot,\rm init} \simeq 7$--$8$~kpc, which includes the present Solar radius; at smaller birth radii its distribution lies above the meteoritic value, indicating that a Sun born there would have inherited a higher ratio than observed. Model~3 behaves similarly but is shifted upward: it is consistent with the meteoritic value only at $R_{\odot,\rm init} = 8$~kpc and exceeds it at all smaller radii. Model~4, in which the GCR production is suppressed in dense gas, reaches the meteoritic value only at $R_{\odot,\rm init} = 2$~kpc and falls below it elsewhere.

These results directly address the long-standing question of whether the early Solar System's $^{10}$Be/$^{9}$Be could have been acquired by inheritance from the parent molecular cloud alone, or whether a late injection or local irradiation was required. In all four models there exists a range of birth radii for which the inherited ratio at star particle formation is consistent with the meteoritic value. Even in Model~4, which imposes a deliberately strong suppression of the GCR-induced $^{10}$Be production rate in dense gas, the inherited ratio falls short of the meteoritic value by only a factor of a few outside the innermost disc---rather than the much larger deficit that would leave a localized irradiation as the dominant source. This suggests that inheritance of GCR-produced $^{10}$Be from the parent molecular cloud can account for the early Solar System's $^{10}$Be/$^{9}$Be without invoking a late injection or local irradiation.

Figure~\ref{fig:initial_10Be_9Be_solar_birth_radius} can be used to address two distinct questions, which require different diagnostics. The first question is where the Sun was born: for this, the percentile criterion introduced above (consistency of the meteoritic value with the 25th--75th percentile range of the stellar distribution) identifies the birth radii for which the inherited ratio at star particle formation is consistent with the meteoritic value, and is the diagnostic adopted here when considering the Solar birth radius. The second, more basic question is whether GCR irradiation supplies enough $^{10}$Be at all: for this, a useful complementary diagnostic is one-sided: because any closed-box evolution between isolation of the protosolar material and CAI closure lowers the ratio through radioactive decay (Section~\ref{subsec:limitations}), birth radii at which the median inherited ratio lies at or above the meteoritic value are not excluded on the grounds of an insufficient inherited $^{10}$Be abundance. An inner-galaxy birth is therefore compatible with a sufficient inherited abundance of GCR-produced $^{10}$Be in Models~1--3, although the radius at which the typical inherited abundance becomes sufficient is itself model-dependent (the median inherited ratio lies at or above the meteoritic value out to $R_{\odot,\rm init} \simeq 8$, 7, and 4~kpc in Models~2, 3, and 1, respectively, and only at $\simeq 2$~kpc in Model~4).

Our results should be contrasted with those of \citet{TatischeffEtAl2014}, whose local production rate and $^{9}$Be abundance our normalization follows. From the steady-state irradiation of the presolar molecular cloud by the mean galactic CR population (the CR ``sea''), they obtained $^{10}$Be/$^{9}$Be $= P_{\mathrm{^{10}Be}}\,\tau_{\mathrm{^{10}Be}}\,/\,(^{9}\mathrm{Be/H}) \simeq 9\times10^{-5}$ at the present epoch, where $\tau_{\mathrm{^{10}Be}} = t_{1/2}/\ln 2 \simeq 2.0$~Myr, rising to $1.3\times10^{-4}$ at the time of Solar System formation with their adopted GCR enhancement factor of 1.5---the analogue of our $\Psi$. This fell short of the FUN-CAI ratio of $\sim 3\times10^{-4}$ that was their reference value by a factor of $\sim 2.3$ (and lies a factor of $\sim 5$ below the value of $\sim 7\times10^{-4}$ adopted here), leading them to conclude that the presolar cloud must have been irradiated by an additional source of CRs: either CRs from $\sim 50$ supernovae in a superbubble blown by a rare, $\gtrsim 20{,}000$-member star cluster, or freshly accelerated CRs escaping from a single nearby SNR with a specific expansion history. Our simulations reproduce their starting point: the volume-weighted mean ratio of the gas at $R = 8$~kpc in Model~2, $\sim 1\times10^{-4}$ (Figure~\ref{fig:radial_profile_10Be_9Be}, left), agrees with their steady-state CR-sea estimate, providing a consistency check on our normalization. The picture changes, however, for the material that actually forms stars: the median ratio inherited by stars forming at $R = 8$~kpc in Model~2 is $\sim 5\times10^{-4}$ (Figure~\ref{fig:radial_profile_10Be_9Be}, right), a factor of $\sim 5$ above the CR-sea value at the same radius, because CRs are injected by SN feedback where stars form, so that the natal environments of stars are systematically CR-enriched relative to the mean ISM (Section~\ref{subsec:10Be_9Be_distribution}). In other words, the additional CR source inferred by \citet{TatischeffEtAl2014} need not be a rare superbubble or a fine-tuned nearby SNR: at the population level, it is supplied by the ordinary spatial correlation between star formation and CR injection, which a one-zone estimate based on the mean ISM cannot capture by construction; resolving it is what the galactic-scale simulation adds. Consistent with this picture, the $^{10}$Be produced inside individual SNRs contributes only a minor fraction of the stellar inventory in our simulations (Section~\ref{subsec:contribution_SNR_10Be}; see also Sections~\ref{subsec:10Be_distribution_in_the_galactic_disc} and \ref{subsec:model_dependence_10Be_distribution}). It is this environmental enhancement, combined with the assumed higher star formation rate at the epoch of Solar System formation, that allows the inherited ratio to reach the meteoritic value.

It is instructive to compare this picture, based on the whole population of newly formed stars rather than a single presolar cloud, with a previous galactic-scale study of the stellar-produced SLRs $^{26}$Al and $^{60}$Fe \citep{FujimotoEtAl2018}. A common feature of the two studies is that star formation does not sample the ISM randomly: it is spatially correlated with the processes responsible for radionuclide production, so that newly formed stars preferentially sample environments with enhanced SLR abundances. The resulting distributions, however, differ markedly. In \citet{FujimotoEtAl2018}, the $^{26}$Al/$^{27}$Al and $^{60}$Fe/$^{56}$Fe ratios inherited by newly formed stars showed broad distributions spanning orders of magnitude at a given radius, while varying only weakly with galactocentric radius over $R = 4$--$10$~kpc (their Appendix~A). By contrast, the inherited $^{10}$Be/$^{9}$Be ratio in the present simulations has a comparatively narrow distribution at a given radius and exhibits a systematic radial trend that follows the large-scale CR field (Figure~\ref{fig:radial_profile_10Be_9Be}, right). This difference reflects the distinct production and transport of these SLRs: $^{26}$Al and $^{60}$Fe retain a strong sensitivity to the local, stochastic history of massive-star feedback, whereas GCR-produced $^{10}$Be reflects the spatially smoother CR environment (Section~\ref{subsec:10Be_9Be_distribution}). Different SLRs therefore encode complementary spatial aspects of the Solar birth environment, from the local enrichment history to the larger-scale galactic CR field.

The implications for the Solar birth radius, on the other hand, are necessarily weaker and depend strongly on the assumed model. A birth radius close to the present Solar radius would require little or no radial migration, whereas an inner-disc origin would imply substantial outward migration \citep[see][and references therein]{FujimotoInutsukaBaba2023}. We therefore conclude that, while inheritance of GCR-produced $^{10}$Be can account for the meteoritic $^{10}$Be/$^{9}$Be in the models explored here, the ratio by itself does not place a robust constraint on the Solar birth radius, which remains strongly model-dependent. We note, nonetheless, that the smaller birth radii allowed by some of our models would be consistent with the long-standing suggestion, based on Galactic metallicity gradients and radial migration, that the Sun is more metal-rich than typical stars of similar age at $R \sim 8$~kpc and may therefore have originated at a smaller galactocentric radius before migrating outward; published estimates along these lines span $\sim4$ to $\sim7$~kpc \citep[e.g.,][]{WielenFuchsDettbarn1996,FrankelEtAl2018,MinchevEtAl2018,BabaSaitohTsujimoto2023,LuEtAl2024}. Whether such migration is actually required is, however, debated. High-precision abundances of nearby solar-type stars reveal two distinct sequences in the age--metallicity relation, with little scatter within each; the Sun lies on the younger sequence, and the small intrinsic scatter reduces the need for radial migration to explain the dispersion in [Fe/H] at a given age \citep{NissenEtAl2020}. These sequences are reproduced by two-infall models of Galactic chemical evolution \citep{ChiappiniMatteucciGratton1997,SpitoniEtAl2019} without invoking migration. An inner-disc origin of the Sun is thus a possibility consistent with our models rather than a requirement, and our results do not by themselves discriminate between these views.

Finally, we do not apply a separate correction for metallicity evolution over the past 4.6~Gyr. On the $^{9}$Be side, no correction is required: the adopted normalization is the Solar abundance, which is by construction the abundance of the proto-Solar material. On the $^{10}$Be side, the production of thermal $^{10}$Be retained in molecular clouds is dominated by GCR protons and $\alpha$-particles impinging on ambient CNO nuclei \citep{TatischeffEtAl2014}, and is therefore proportional to the CNO abundance of the ambient gas. Given the substantial uncertainty in the chemical evolution of the Galactic disc over this interval \citep[e.g.,][]{Rebassa-MansergasEtAl2016,DonerAkOnal2023}, we do not impose an additional metallicity scaling and instead treat its possible effect as a systematic uncertainty.

\subsection{Inherited \texorpdfstring{$^{10}$Be/$^{9}$Be}{10Be/9Be} across the Galactic disc at the epoch of Solar System formation}
\label{subsec:inherited_10Be_9Be_across_disc}

We now turn to a related but distinct question: what $^{10}$Be/$^{9}$Be ratios did star-forming regions \emph{elsewhere} in the Galactic disc inherit at the epoch of Solar System formation? Answering this requires the radial profile of $^{9}$Be/H across the disc at that epoch, which is not observationally known: Be abundances have been measured only in nearby stars, and no spatially resolved Be/H gradient across the Galactic disc is available. Two indirect anchors are available. Observationally, the stellar Be/H abundance closely tracks Fe/H with a slope near unity \citep[e.g.,][]{SmiljanicEtAl2009, BoesgaardEtAl2011}, so the $^{9}$Be gradient is expected to follow the observed metallicity gradient of the disc, $\simeq -0.04$ to $-0.06$~dex~kpc$^{-1}$ \citep[e.g.,][]{GenovaliEtAl2014, ArellanoCordovaEtAl2020}. Theoretically, the galactic chemical evolution model of \citet{Prantzos2012}, which provided the first explicit prediction of the radial Be profile of the disc, yields a steeper gradient of $-0.085$~dex~kpc$^{-1}$ at the present day and $\sim -0.1$~dex~kpc$^{-1}$ at the epoch of Solar System formation.

Guided by these anchors, we construct the $^{9}$Be/H profile as
\begin{equation}
\label{eq:9Be_gradient_R_init}
(^{9}{\rm Be}/{\rm H})(R) = (^{9}{\rm Be}/{\rm H})_{\odot} \times 10^{\alpha (R - R_{\odot,\rm init})},
\end{equation}
anchored to the protosolar value at the assumed Solar birth radius $R_{\odot,\rm init}$, and explore $\alpha = -0.05$ and $-0.1$~dex~kpc$^{-1}$, bracketing the metallicity-anchored and GCE-motivated values, for $R_{\odot,\rm init} = 8$, $6$, $4$, and $2$~kpc. The $^{10}$Be of each star particle is again multiplied by $\Psi = 2.0$. For illustration we show Model~2, the intermediate-diffusivity case. We emphasize that the resulting profiles are \emph{conditional}: they are normalized to the protosolar $^{9}$Be abundance at the assumed Solar birth radius, and should not be regarded as a unique chemical-evolution prediction for the Galactic disc at that epoch.

\begin{figure}
\centering
\includegraphics[width=0.98\columnwidth]{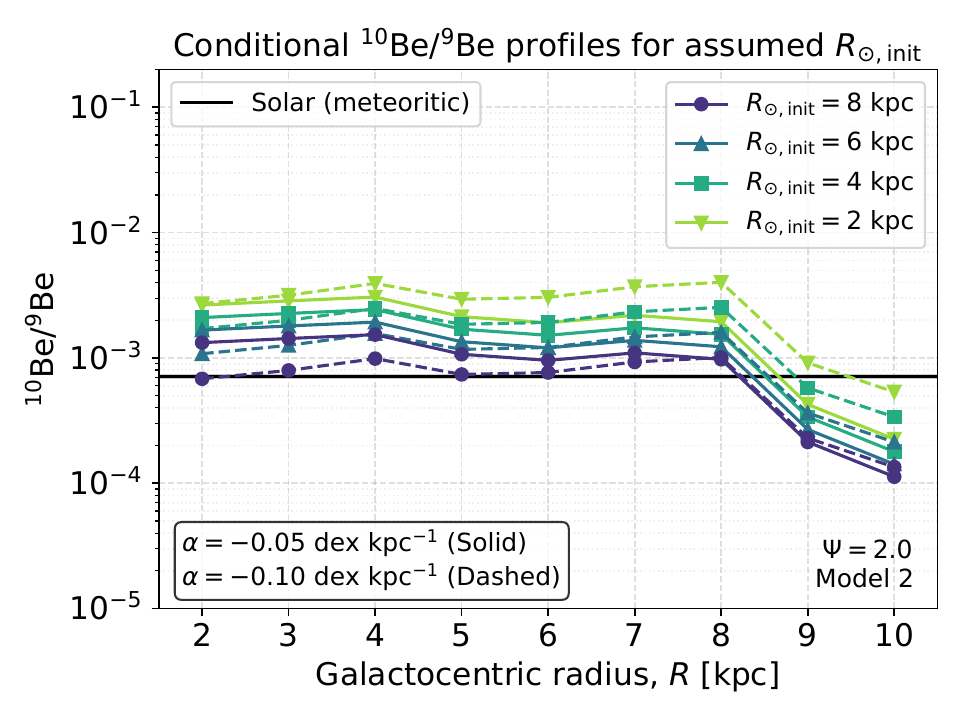}
\caption{
Radial profiles of the $^{10}$Be/$^{9}$Be ratio inherited by newly formed stars at the epoch of Solar System formation, computed for Model~2, the intermediate-diffusivity case with an SFR enhancement factor $\Psi = 2.0$ and a $^{9}$Be/H radial gradient anchored to the protosolar value at the assumed Solar birth radius $R_{\odot,\rm init}$ (Equation~\ref{eq:9Be_gradient_R_init}). Colours show $R_{\odot,\rm init} = 8$, $6$, $4$, and $2$~kpc; solid and dashed lines show median profiles for $\alpha = -0.05$ and $-0.10$~dex~kpc$^{-1}$, respectively. The horizontal black line marks the initial Solar System value of $\sim 7\times10^{-4}$ inferred from meteoritic inclusions \citep{DunhamEtAl2022}. These are conditional profiles normalized at $R_{\odot,\rm init}$, not a chemical-evolution prediction for the disc (see text).
}
\label{fig:radial_profile_10Be_9Be_solar_epoch}
\end{figure}

Figure~\ref{fig:radial_profile_10Be_9Be_solar_epoch} shows the resulting profiles. Their behaviour follows directly from the construction: a smaller $R_{\odot,\rm init}$ assigns a lower $^{9}$Be/H to the entire disc and shifts the whole profile upward, while a steeper gradient (larger $|\alpha|$) raises $^{9}$Be/H---and hence lowers the ratio---inside the anchor radius and does the opposite outside it. The most notable feature, however, is common to all cases: the profiles are remarkably flat over $R \simeq 2$--$8$~kpc, at or above the meteoritic value, before dropping at $R \gtrsim 9$~kpc. The flatness arises because the intrinsic radial decline of the simulated $^{10}$Be/$^{9}$Be (Section~\ref{subsec:10Be_9Be_distribution}) is largely compensated by the assumed negative $^{9}$Be gradient.

The implication is that, in Model~2, an initial $^{10}$Be/$^{9}$Be at or above the meteoritic value was not peculiar to the Solar birth environment: newly formed stars across most of the star-forming disc, $R \simeq 2$--$8$~kpc, would have inherited comparable or higher ratios, regardless of where within this range the Sun itself formed. This is consistent with the conclusion of Section~\ref{subsec:initial_10Be_9Be_solar_birth_environment} that no late injection or local irradiation is required. This inference is subject to two caveats. First, it is based on Model~2; we have verified that a negative radial $^{9}$Be gradient similarly reduces the radial variation of the inherited ratio in the other models, but the level relative to the meteoritic value remains model-dependent. For the plausible diffusion cases represented by Models~1--3, and depending on the assumed Solar birth radius and $^{9}$Be gradient, inherited ratios comparable to or exceeding the meteoritic value can extend over a substantial fraction of the star-forming disc, whereas in Model~4, with its deliberately strong suppression of production in dense gas, they fall short at almost all radii, reaching the meteoritic value only for the most favourable combination of $R_{\odot,\rm init} = 2$~kpc and $\alpha = -0.1$. Second, it does not extend to the outer disc, $R \gtrsim 9$~kpc, where the inherited ratios fall below the meteoritic value in all cases considered.

The same scaling implies a dependence on the epoch of star formation. Because the inherited $^{10}$Be/$^{9}$Be scales with the CR energy density and hence, to first order, with the Galactic star formation rate (Section~\ref{subsec:initial_10Be_9Be_solar_birth_environment}), stars forming today, when the star formation rate is lower than at the epoch of Solar System formation, would inherit ratios lower than the Sun's by roughly the factor $\Psi$, whereas stars formed at earlier epochs of more vigorous star formation could have inherited substantially higher ratios, further enhanced by the lower $^{9}$Be/H of the less enriched early ISM. We note, however, that $^{10}$Be is of interest as a tracer of the CR environment rather than as a heat source for protoplanetary discs: at the meteoritic ratio, $^{10}$Be/H $\approx 2\times10^{-14}$, four orders of magnitude below $^{26}$Al/H $\approx 1.5\times10^{-10}$ at the canonical $^{26}$Al/$^{27}$Al $\approx 5\times10^{-5}$, and its decay deposits only $\sim 0.2$~MeV per decay compared with $\sim 3$~MeV for $^{26}$Al, so that its heating rate is $\sim 10^{-5}$ that of $^{26}$Al. Moreover, the region-to-region variation of $^{26}$Al, which spans orders of magnitude \citep{FujimotoEtAl2018}, far exceeds the factor-of-a-few epoch-to-epoch variation of $^{10}$Be, so that the heating of protoplanetary discs is set by the local star-forming environment rather than by the Galactic epoch.

\subsection{Contribution of SNR-produced \texorpdfstring{$^{10}$Be}{10Be}}
\label{subsec:contribution_SNR_10Be}

\begin{figure}
\centering
\includegraphics[width=0.98\columnwidth]{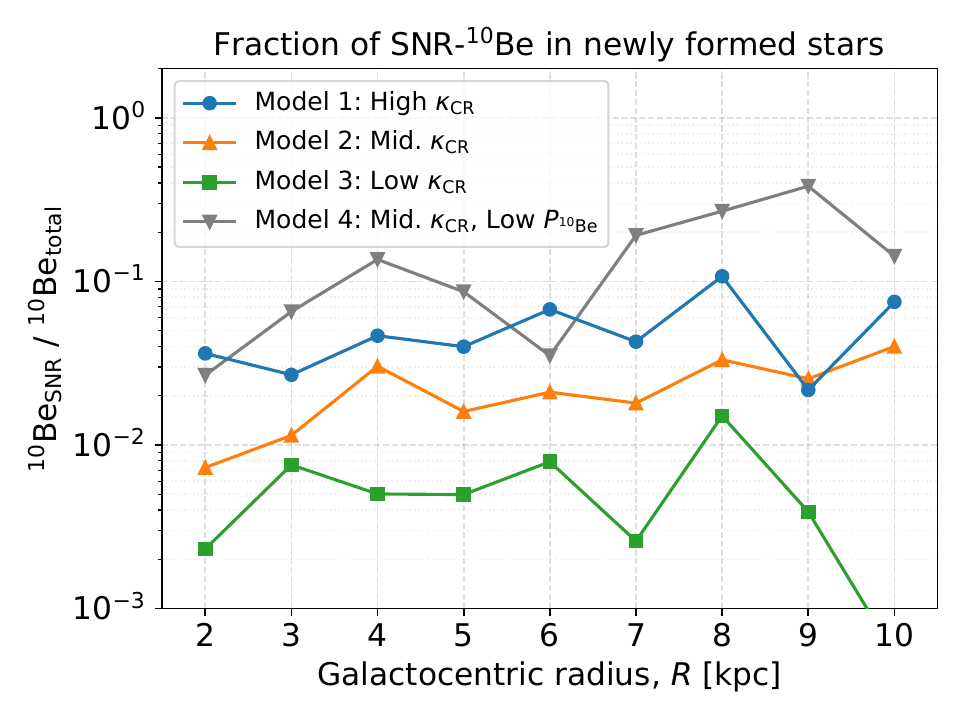}
\caption{
Median fraction of SNR-produced $^{10}$Be in the total $^{10}$Be inherited by newly formed stars, as a function of galactocentric radius at $t = 600$~Myr. Stars are selected within radial bins of 1~kpc, and the stellar sample comprises stars formed within the last 5~Myr. The four models are represented by the same colours and markers as in Figure~\ref{fig:initial_10Be_9Be_solar_birth_radius}.
}
\label{fig:radial_profile_fraction_SNR_10Be_stars}
\end{figure}

In Section~\ref{subsec:initial_10Be_9Be_solar_birth_environment} we argued that the early Solar System's $^{10}$Be/$^{9}$Be can be accounted for by the inheritance of GCR-produced $^{10}$Be. Here we quantify the contribution of SNR-produced $^{10}$Be to the $^{10}$Be inherited by newly formed stars. Figure~\ref{fig:radial_profile_fraction_SNR_10Be_stars} shows its median fraction as a function of galactocentric radius. The ordering of the models follows the production physics discussed in Section~\ref{subsec:model_dependence_10Be_distribution}: the SNR fraction is lowest in Model~3, which has the highest CR energy density, and highest in Model~4, in which the GCR production is suppressed in dense gas.

The stellar SNR fractions are somewhat higher than the gas mass-weighted values in Figure~\ref{fig:radial_profile_model_comparison} (bottom right), as expected: newly formed stars trace the actively star-forming gas most directly, which is also where recent supernovae inject SNR-produced $^{10}$Be. Although the fraction fluctuates with radius, the median is around ten percent for Model~4, several percent for Model~1, a few percent for Model~2, and just below one percent for Model~3. Across all four models, therefore, the $^{10}$Be inherited by newly formed stars is dominated by the GCR-produced component, and the inheritance scenario of Section~\ref{subsec:initial_10Be_9Be_solar_birth_environment} does not require a direct injection of SNR-produced $^{10}$Be.

\subsection{Limitations}
\label{subsec:limitations}

Four limitations of our analysis deserve explicit mention. They affect the detailed comparison with the meteoritic value---and, in particular, the results for the inner disc---but not the main conclusion of this paper.

The first concerns the gap between the formation of a star particle in the simulation and the formation of CAIs in a real protoplanetary system. Each star particle inherits the $^{10}$Be/$^{9}$Be ratio of its natal gas cell at the moment of its formation, at the $10$~pc resolution of the simulation. Because star particles form stochastically rather than immediately upon crossing the density threshold (Section~\ref{subsec:star_formation_and_feedback}), and because the production and decay of $^{10}$Be are followed on the fly in the gas, the finite delay between the onset of gravitational instability and final collapse is already included in the inherited ratio \citep[see also][]{FujimotoEtAl2018}. What remains unresolved is the final stage of protostellar collapse below our resolution scale: once the protosolar material becomes isolated from the GCR-irradiated ISM (a ``closed box''), its $^{10}$Be only decays until it is recorded in CAIs, by a factor of two per half-life ($t_{1/2} = 1.387$~Myr), so that an order-of-magnitude reduction would require $\simeq 4.6$~Myr of isolation. Since the epoch of this isolation is uncertain, we do not apply a decay correction. The net effect nevertheless acts in one direction only, reducing the ratio recorded in CAIs relative to that inherited at star particle formation. The tendency of Models~1--3 to inherit ratios at or above the meteoritic value at sufficiently small birth radii (Figure~\ref{fig:initial_10Be_9Be_solar_birth_radius}) can therefore accommodate some reduction before CAI closure, although the allowed reduction is model- and radius-dependent: an order-of-magnitude excess, such as that in the innermost disc of Model~3, would require a long isolation interval and may also reflect the possible overestimate of the inner-disc CR energy density discussed below. Model~4 probes only the reduced late-stage production, by suppressing GCR-induced $^{10}$Be production in resolved dense gas; it does not model the isolation-to-CAI decay interval.

The second concerns our treatment of the CRs. The two-fluid model characterizes the entire CR population by a single energy density, carried predominantly by $\sim 1$--$10$~GeV particles, whereas the spallation reactions that produce $^{10}$Be are governed by the number flux of sub-GeV CRs. We therefore assume that the shape of the GCR spectrum does not vary significantly across the disc, so that the $^{10}$Be production rate scales with the total CR energy density (Equation~\ref{eq:P_10Be}), with its normalization set by the locally observed GCR fluxes and CR energy density (Section~\ref{subsec:10Be_production_model}). Because CR transport and energy losses are energy dependent, this assumption cannot hold exactly; the parameter study in $\kappa_{\rm CR}$, whose range spans the diffusivities plausible for the sub-GeV CRs relevant to $^{10}$Be production (Section~\ref{subsec:parameter_study}), serves as a sensitivity test of its consequences. We also neglect CR energy losses due to Coulomb collisions, ionization of the ambient gas, and hadronic interactions. For the $\sim$GeV CRs that carry the bulk of the CR energy density these losses are modest: the hadronic loss time, $t_{\mathrm{had}} \simeq 40\ \mathrm{Myr}\,(n_{\mathrm{H}}/1\ \mathrm{cm^{-3}})^{-1}$ \citep[e.g.,][]{PfrommerEtAl2017}, is long compared with the time to diffuse across a dense structure, $t_{\mathrm{diff}} \sim L^{2}/(6\kappa_{\rm CR}) \simeq 13\ \mathrm{kyr}\,(L/50\ \mathrm{pc})^{2}\,(\kappa_{\rm CR}/10^{28}\ \mathrm{cm^{2}\ s^{-1}})^{-1}$, so that even at $n_{\mathrm{H}} = 100\ \mathrm{cm^{-3}}$ a CR loses only a few percent of its energy per crossing, although the losses may become non-negligible in our lowest-diffusivity model or if CR transport is suppressed within clouds \citep{SkillingStrong1976, EverettZweibel2011}. For the sub-GeV CRs that dominate $^{10}$Be production, however, ionization losses are far more severe and penetration into dense clouds may be strongly suppressed; because the two-fluid model does not evolve the CR spectrum, this effect is subsumed in the assumption stated above. Model~4, in which $P_{\mathrm{^{10}Be}}$ is reduced by a factor of ten for $n_{\mathrm{H}} > 100\ \mathrm{cm^{-3}}$, tests the sensitivity of the inherited $^{10}$Be to such a suppression in the star-forming gas, but does not model the energy losses themselves or the subsequent transport of the attenuated CR population. A self-consistent treatment of CR energy losses \citep[e.g., using the loss terms of][]{PfrommerEtAl2017} is left to future work.

The third concerns the CR energy density in the inner disc. Our simulated $\epsilon_{\rm CR}$ declines steeply with radius, with fitted slopes of $\simeq -0.15$ to $-0.21$~dex~kpc$^{-1}$ over $R = 2$--$10$~kpc depending on the CR diffusion coefficient (Section~\ref{subsec:model_dependence_10Be_distribution}), so that, although all four models reproduce the locally observed $\sim 1$~eV~cm$^{-3}$ at the Solar radius, $\epsilon_{\rm CR}$ at $R \simeq 2$--$3$~kpc reaches $\sim 5$~eV~cm$^{-3}$ in Model~1 and $\sim 10$--$30$~eV~cm$^{-3}$ in Models~2--4 (Figure~\ref{fig:radial_profile_model_comparison}). By contrast, the GeV CR density inferred from diffuse gamma-ray observations exceeds the local value by only a factor of a few in the inner Galaxy \citep[][see also \citealt{ShimodaAsano2024}]{AckermannEtAl2012, AceroEtAl2016, YangAharonianEvoli2016}---the long-standing CR gradient problem \citep[e.g.,][]{StrongEtAl2004}---and our simulated gamma-ray longitude profiles are correspondingly more centrally concentrated than the EGRET observations, with a central intensity comparable to the observed one in Model~1 but exceeding it by factors of $\simeq 1.5$ and $\simeq 3$ in Models~2 and 3 (Appendix~\ref{sec:galactic_gamma_ray_longitude_profile}). These constraints apply to CRs at $\gtrsim$~GeV energies: Model~1, whose diffusivity is the canonical GeV value, lies within a factor of $\sim 2$ of the inferred inner-disc values, whereas the larger excesses arise in Models~2--4, whose lower diffusivities are motivated by the sub-GeV CRs that dominate $^{10}$Be production (Section~\ref{subsec:parameter_study}) and whose galactic-scale distribution is not directly probed by the gamma-ray data. We do not attempt to resolve this discrepancy here, but note two consequences. First, the inherited $^{10}$Be/$^{9}$Be ratios in the inner disc should be regarded as upper limits: a flatter true CR profile would lower them, so that the statement in Section~\ref{subsec:inherited_10Be_9Be_across_disc} that the inherited ratio lies at or above the meteoritic value over $R \simeq 2$--$8$~kpc should be read as ``at or near''; the spread among Models~1--3 already brackets part of this uncertainty, since a higher diffusivity yields both a lower and a somewhat flatter CR profile. The same caution applies to the inner birth radii in Figure~\ref{fig:initial_10Be_9Be_solar_birth_radius}, which, if anything, makes small birth radii more rather than less compatible with the meteoritic value and further weakens the constraint on the Solar birth radius. Second, even if the simulated inner-disc CR density were realistic, the same spallation would have enhanced the $^{9}$Be abundance there over Galactic history, so that our assumption of a spatially uniform $^{9}$Be/H would underestimate $^{9}$Be in the inner disc and again overestimate $^{10}$Be/$^{9}$Be. Either way, the inner-disc ratios are biased upward; the negative $^{9}$Be gradients explored in Section~\ref{subsec:inherited_10Be_9Be_across_disc} act in this direction and partially capture the effect. 

The fourth is that we neglect CR heating of the gas, via both Alfv\'en-wave damping and Coulomb collisions. Such heating can help drive galactic winds and fountains \citep[e.g.,][]{ShimodaInutsuka2022, ShimodaAsanoInutsuka2026}, which would redistribute both $^{9}$Be and $^{10}$Be between the disc and the halo and reduce the gas and CR content retained in the disc. A quantitative characterization of the outflows in our simulations is beyond the scope of this work, but we note that the neglect of this heating acts in the same direction as the inner-disc excess discussed above.

These limitations primarily affect the quantitative radial distribution of the inherited $^{10}$Be/$^{9}$Be ratio and hence the inferred Solar birth radius, with the inner-disc predictions being particularly uncertain. They therefore reinforce our conclusion that the Solar birth radius cannot be robustly constrained from $^{10}$Be/$^{9}$Be alone. The more general result that GCR-produced $^{10}$Be dominates the $^{10}$Be inventory inherited by newly formed stars nevertheless remains robust within the models explored here.

\section{Conclusions}
\label{sec:conclusions}
 
We have performed three-dimensional hydrodynamic simulations of a Milky Way--like disc that self-consistently evolve the multi-phase interstellar medium and cosmic rays, together with the production, transport, and decay of both GCR-produced and SNR-produced $^{10}$Be. We evolved four models that differ in the CR diffusion coefficient $\kappa_{\rm CR}$ and in the GCR production rate of $^{10}$Be in dense gas, with the three values of $\kappa_{\rm CR}$ spanning a plausible range for the low-energy CRs relevant to $^{10}$Be production. By tracing the $^{10}$Be/$^{9}$Be ratio inherited by newly formed stars as a function of galactocentric radius, we have examined whether the high initial $^{10}$Be/$^{9}$Be ratio of the early Solar System can be explained by the inheritance of galactic $^{10}$Be, and what this implies for the Sun's birth environment. Our main findings are as follows.
 
\begin{enumerate}
\item Nearly all of the $^{10}$Be in the disc is produced by GCR interactions with the interstellar gas (Figure~\ref{fig:zoom_projections}, bottom row; Figure~\ref{fig:radial_profile_model_comparison}, bottom right). SNR-produced $^{10}$Be plays only a secondary role, contributing from below one percent to about ten percent of the $^{10}$Be inherited by newly formed stars, depending on the CR transport model (Figure~\ref{fig:radial_profile_fraction_SNR_10Be_stars}).
\item The total $^{10}$Be density inherits the clumpy, high-contrast morphology of the gas (Figure~\ref{fig:zoom_projections}, top row), but its radial decline is steeper than that of either the gas or the CR field alone (Figure~\ref{fig:radial_profile_model_2}), because the GCR production rate of $^{10}$Be is proportional to the product of the gas density and the CR energy density.
\item The $^{10}$Be/$^{9}$Be abundance ratio, in contrast, is governed almost entirely by the CR energy density: it shows a smooth spatial morphology (Figure~\ref{fig:zoom_projections}, bottom right) and a clear radial decline (Figure~\ref{fig:radial_profile_10Be_9Be}, left) that mirror the CR field, with little sensitivity to the clumpy structure of the gas.
\item The $^{10}$Be/$^{9}$Be ratio inherited by newly formed stars decreases by more than an order of magnitude between $R = 2$ and $10$~kpc, mirroring the radial trend of the volume-weighted gas ratio (Figure~\ref{fig:radial_profile_10Be_9Be}, right). The stellar median lies above the gas value at the same radius (by up to a factor of $\sim 5$), reflecting the enhanced CR energy density in the dense, star-forming regions that newly formed stars trace directly. This environmental enhancement is the key difference from previous one-zone estimates based on the mean ISM (e.g., Tatischeff et al. 2014): our volume-weighted gas ratio at the Solar radius reproduces their CR-sea value of $\sim 1\times10^{-4}$, but because stars form where supernovae inject CRs, their natal gas is systematically CR-enriched, so that the additional CR source invoked by one-zone models is supplied by the ordinary spatial correlation between star formation and CR injection.
\item Accounting for a possibly higher star formation rate at the epoch of Solar System formation 4.6~Gyr ago---an SFR enhancement factor $\Psi = 2.0$, assumed to raise the CR energy density and hence the $^{10}$Be production by the same factor---the inheritance of GCR-produced $^{10}$Be from the parent molecular cloud can account for the early Solar System's $^{10}$Be/$^{9}$Be ratio in the models explored here, without requiring a late injection or local irradiation (Figure~\ref{fig:initial_10Be_9Be_solar_birth_radius}). Even in Model~4, which imposes a deliberately strong suppression of GCR-induced $^{10}$Be production in dense gas, the inherited ratio reaches the meteoritic value in the innermost disc and falls short of it by only a factor of a few elsewhere, rather than the much larger deficit that would leave a localized irradiation as the dominant source.
\item Allowing in addition for a plausible negative radial gradient of the $^{9}$Be abundance, anchored to the protosolar value at the assumed birth radius, largely compensates the intrinsic radial decline of the inherited $^{10}$Be/$^{9}$Be, yielding nearly flat profiles over $R \simeq 2$--$8$~kpc (Figure~\ref{fig:radial_profile_10Be_9Be_solar_epoch}, shown for Model~2; the other models behave similarly), although the absolute normalization relative to the meteoritic value remains model-dependent. For the plausible diffusion cases represented by Models~1--3, inherited ratios comparable to or exceeding the meteoritic value can extend over a substantial fraction of the disc, so that the early Solar System's $^{10}$Be/$^{9}$Be need not represent an unusually high value among contemporaneous star-forming regions.
\item The birth radii consistent with the meteoritic value depend strongly on the adopted CR diffusivity (Figure~\ref{fig:initial_10Be_9Be_solar_birth_radius}): Model~1 is consistent with $R_{\odot,\rm init} \simeq 3$--$5$~kpc (and, given its broad scatter, also with $\simeq 8$~kpc), Model~2 with $\simeq 7$--$8$~kpc, and Model~3 with $\simeq 8$~kpc, while Model~4 is consistent only at $\simeq 2$~kpc. The $^{10}$Be/$^{9}$Be ratio alone therefore does not tightly constrain the Solar birth radius, although birth radii smaller than the present Solar radius would be consistent with, though not required by, arguments based on Galactic metallicity gradients that the Sun originated at a smaller galactocentric radius and subsequently migrated outward (Section~\ref{subsec:initial_10Be_9Be_solar_birth_environment}). The systematic radial dependence of the simulated inherited $^{10}$Be/$^{9}$Be stands in contrast to the weak radial dependence found for the stellar-produced SLRs $^{26}$Al and $^{60}$Fe \citep{FujimotoEtAl2018}, highlighting that \emph{different SLRs encode complementary spatial aspects of the Solar birth environment}. The sensitivity of the inferred radius to CR diffusivity suggests that improved modelling of low-energy CR transport in the multiphase ISM could make ${}^{10}$Be/${}^{9}$Be a more quantitative tracer of the Solar birth environment.
\end{enumerate}

Taken together, our results indicate that, although the Solar birth radius remains model-dependent and the relevant sub-GeV CR diffusivity remains uncertain, within the range of CR transport models explored here the meteoritic $^{10}$Be/$^{9}$Be $\approx 7\times10^{-4}$ can arise naturally from the Galactic environment of the Sun's parent molecular cloud. The inherited $^{10}$Be is dominated by GCR spallation in the ISM, enhanced in the star-forming gas relative to the mean ISM by the spatial correlation between star formation and CR injection, with SNR-produced $^{10}$Be contributing at most about ten percent. Additional local sources of $^{10}$Be, such as irradiation of the protoplanetary disc by energetic particles from the young Sun, are therefore not required to explain the meteoritic abundance.

\begin{acknowledgments}

We thank the anonymous referee for constructive comments that improved the manuscript.
This work was supported by JSPS KAKENHI Grant Number 25K24687. 
Numerical computations and analyses were carried out on Cray XC50, HPE Cray XD2000, and the analysis servers at the Center for Computational Astrophysics, National Astronomical Observatory of Japan.
Large language models (Claude, ChatGPT, and Gemini) were used to assist with language editing and shortening of the text; all scientific content and conclusions were written, verified, and are the responsibility of the authors.

\end{acknowledgments}

\begin{contribution}

Y.F. conceived the project, performed the simulations and analysis, and wrote the manuscript. S.J.D., J.S., and S.I. contributed to the physical interpretation of the results and edited the manuscript.


\end{contribution}

%

\software{Enzo \citep{BryanEtAl2014},  
          SLUG \citep{KrumholzEtAl2015}, 
          Grackle \citep{SmithEtAl2017},
          yt \citep{TurkEtAl2011},
          NumPy \citep{HarrisEtAl2020},
          Matplotlib \citep{Hunter2007}
          }


\appendix

\section{Comparison with the Galactic gamma-ray longitude profile}
\label{sec:galactic_gamma_ray_longitude_profile}

\begin{figure}
\centering
\includegraphics[width=0.5\columnwidth]{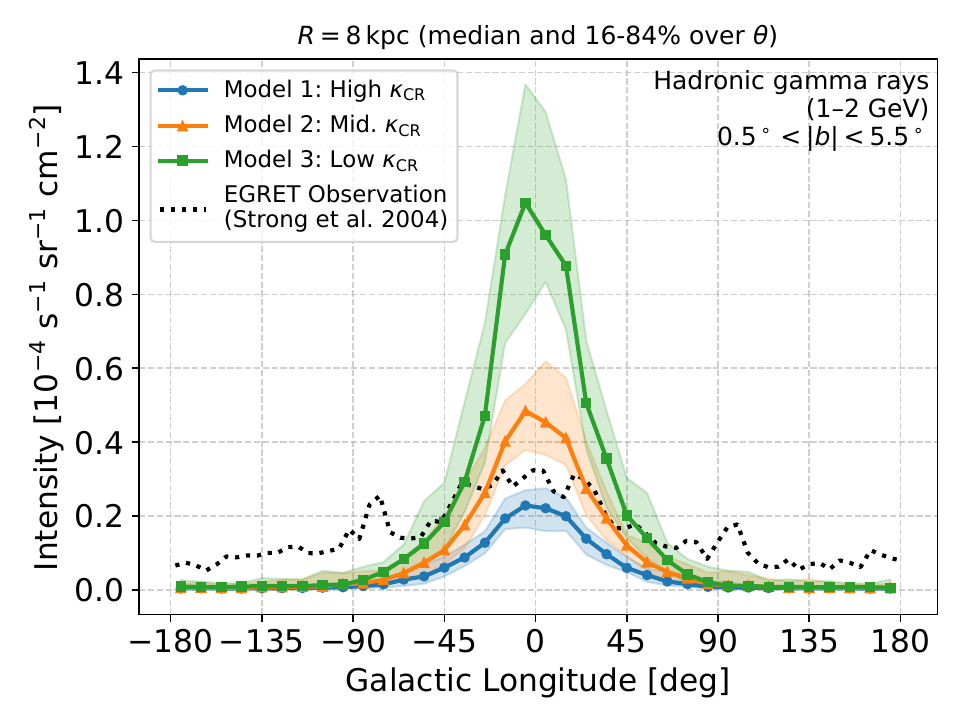}
\caption{Galactic longitude profile of the simulated 1--2~GeV hadronic gamma-ray emission for the latitude range $0.5^\circ<|b|<5.5^\circ$, for Models~1--3 at $t=600$~Myr. The observer is placed at $R=8$~kpc in the galactic mid-plane. For each model the profile was calculated for 36 azimuthal positions of the observer sampling $0^\circ$--$360^\circ$ in $10^\circ$ steps; the solid curve shows the median profile and the shaded region the 16th--84th percentile range. The EGRET data from \citet[][their Figure~3, averaged over the same latitude range]{StrongEtAl2004} are shown for comparison.}
\label{fig:longitude_profile_gamma}
\end{figure}

To assess whether the simulated CR distribution is broadly consistent with observations, we constructed the Galactic longitude profile of hadronic gamma-ray emission from the simulations at $t = 600$~Myr for Models~1--3 (Model~4 shares the CR field of Model~2 and is therefore omitted). For each simulation cell we estimated the 1--2 GeV hadronic gamma-ray emissivity per unit volume $j_\gamma$, assuming that the emissivity is proportional to the product of the hydrogen number density $n_{\rm H}$ and the CR energy density $\epsilon_{\rm CR}$,
\begin{equation}
j_\gamma \propto n_{\rm H}\,\epsilon_{\rm CR}.
\end{equation}

The proportionality constant was calibrated using the locally observed hadronic gamma-ray emissivity per H atom $q_\gamma$. Specifically, we adopted a local value of $q_{\gamma,\mathrm{Local}} = 6.5 \times 10^{-28}\ \mathrm{photons\ s^{-1}\ sr^{-1}\ H^{-1}}$ for the 1--2 GeV band based on observational constraints from \citet{AbdoEtAl2009} and \citet{Casandjian2015}, together with the locally measured CR energy density $\epsilon_{\mathrm{CR,Local}} = 1.6 \times 10^{-12}\ \mathrm{erg\ cm^{-3}} \approx 1\ \mathrm{eV\ cm^{-3}}$. The emissivity in each simulation cell was then estimated as
\begin{equation}
j_\gamma =
\left(
\frac{q_{\gamma,\mathrm{Local}}}{\epsilon_{\mathrm{CR,Local}}}
\right)
\,\epsilon_{\rm CR}\,n_{\rm H}.
\end{equation}

An observer was placed at $R = 8$~kpc in the galactic mid-plane, and the gamma-ray intensity was computed as a function of the Galactic longitude by integrating the emissivity along each line of sight over the latitude range $0.5^\circ<|b|<5.5^\circ$, matching the range of the EGRET data. Cells within 200~pc of the observer were excluded to avoid the divergence of the inverse-square distance weighting. The resulting intensity profile varies with the observer's azimuthal position due to the nearby non-axisymmetric structures such as gas clouds and bubbles. To account for these local environmental fluctuations, we repeated the calculation for 36 different observer azimuths $\theta$, sampling the full $0^\circ$--$360^\circ$ range in $10^\circ$ steps. The median longitude profile is shown together with the 16th--84th percentile range (corresponding to $\pm 1\sigma$ for a normal distribution).

Figure~\ref{fig:longitude_profile_gamma} compares the resulting longitude profiles with the EGRET data from \citet{StrongEtAl2004}, from which catalogued point sources have been removed. All three models broadly reproduce the observed shape, with enhanced emission toward the galactic centre and a flat distribution at large longitudes, but all are more centrally concentrated than the observations. The central intensity depends strongly on $\kappa_{\rm CR}$: it is comparable to the observed value in Model~1, exceeds it by a factor of $\simeq 1.5$ in Model~2, and by a factor of $\simeq 3$ in Model~3, reflecting the model dependence of the inner-disc CR energy density shown in Figure~\ref{fig:radial_profile_model_comparison} (top right). Since inverse-Compton emission contributes $\sim 20$ percent of the observed intensity toward the centre \citep{StrongEtAl2004}, the hadronic-only profile of Model~1 in fact matches the observed $\pi^{0}$-decay component closely. At large longitudes, by contrast, all models fall below the observations by a factor of several. Part of this deficit reflects the exclusion of the emission from within 200~pc of the observer, which at $0.5^\circ < |b| < 5.5^\circ$ contributes substantially to the observed intensity in directions away from the Galactic centre, and part reflects the steep radial decline of the simulated CR energy density beyond the Solar radius. Both the central excess in Models~2 and 3 and the deficit at large longitudes are thus manifestations of the steeper-than-observed CR gradient discussed in Section~\ref{subsec:limitations}. One possible contributor to this discrepancy is the neglect of CR heating via Alfv\'en-wave damping and Coulomb collisions, which can help drive galactic winds and fountains \citep[e.g.,][]{ShimodaInutsuka2022,ShimodaAsanoInutsuka2026}. Such outflows could preferentially remove CRs and gas from regions of the disc where the supernova rate is highest, potentially reducing the central CR concentration and flattening its radial profile. Overall, the large-scale CR distribution of Model~1 is consistent with the gamma-ray constraints, whereas the lower-diffusivity models, motivated by sub-GeV CRs not directly probed by these data, overpredict the inner-disc emission.


\bibliography{references}{}
\bibliographystyle{aasjournalv7}



\end{CJK*}
\end{document}